\documentclass[twocolumn]{aastex62}

\usepackage{amsmath}

\received{March 13, 2018}
\accepted{July 11, 2018}
\submitjournal{ApJ}

\defcitealias{Yuan:18}{Paper~I}

\shortauthors{D. Yoon et al.}
\begin{document}

\title{ACTIVE GALACTIC NUCLEUS FEEDBACK IN AN ELLIPTICAL GALAXY WITH THE MOST UPDATED
AGN PHYSICS (II): HIGH-ANGULAR MOMENTUM CASE}

\correspondingauthor{Doosoo Yoon}
\email{yoon; fyuan @shao.ac.cn}

\author[0000-0001-8694-8166]{Doosoo Yoon}
\affiliation{Key Laboratory for Research in Galaxies and Cosmology, Shanghai Astronomical Observatory,
Chinese Academy of Sciences, 80 Nandan Road, Shanghai 200030, China}

\author[0000-0003-3564-6437]{Feng Yuan}
\affiliation{Key Laboratory for Research in Galaxies and Cosmology, Shanghai Astronomical Observatory,
Chinese Academy of Sciences, 80 Nandan Road, Shanghai 200030, China}
\affiliation{University of Chinese Academy of Sciences, 19A Yuquan Road, 100049, Beijing, China}

\author[0000-0003-3886-0383]{Zhao-Ming Gan}
\affiliation{Key Laboratory for Research in Galaxies and Cosmology, Shanghai Astronomical Observatory,
Chinese Academy of Sciences, 80 Nandan Road, Shanghai 200030, China}

\author[0000-0002-6405-9904]{Jeremiah P. Ostriker}
\affiliation{Department of Astronomy, Columbia University, 550 W. 120th Street, New York, NY 10027, USA}

\author[0000-0002-7329-9344]{Ya-Ping Li}
\affiliation{Key Laboratory for Research in Galaxies and Cosmology, Shanghai Astronomical Observatory,
Chinese Academy of Sciences, 80 Nandan Road, Shanghai 200030, China}

\author[0000-0002-5708-5274]{Luca Ciotti}
\affiliation{Department of Physics and Astronomy, University of Bologna, via Piero Gobetti 93/2, 40129 Bologna, Italy}

\begin{abstract}
    This is the second paper of our series of works of studying the effects of active galactic
    nuclei (AGN) feedback on the cosmological evolution of an isolated elliptical galaxy by
    performing two-dimensional high-resolution hydrodynamical numerical simulations. In these
    simulations, the inner boundary  is chosen so that the Bondi radius is resolved. Physical
    processes like star formation, SNe Ia and II are taken into account. Compared to previous
    works, the main improvements is that we adopt the most updated AGN physics, which is described
    in detail in the first paper of this series \citep[Paper I]{Yuan:18}. These improvements
    include the discrimination of the two accretion modes of the central AGN and the most updated
    descriptions of the wind and radiation in the two modes. In \citetalias{Yuan:18}, we consider
    the case that  the specific angular momentum of the gas in the galaxy is very low. In this
    paper, we consider the case that the specific angular momentum of the gas is high. In the
    galactic scale, we adopt the gravitational torques raised due to non-axisymmetric structure
    in the galaxy as the mechanism of  the transfer of angular momentum of gas, as proposed in
    some recent works. Since our simulations are axisymmetric, we make use of a parameterized
    prescription to mimic this  mechanism.  Same as \citetalias{Yuan:18}, we investigate the
    AGN light curve, typical AGN lifetime, growth of the black hole mass, AGN duty-cycle, star
    formation, and the X-ray surface brightness of the galaxy. Special attention is paid to
    the effects of specific angular momentum of the galaxy on these properties. We find that
    some results are qualitatively similar to those shown in \citetalias{Yuan:18}, while some
    results such as star formation and black hole growth do show a significant difference due
    to the mass concentration in the galactic disk as a consequence of galactic rotation.
\end{abstract}


~~~~

\section{Introduction}

It is now believed that most massive galaxies harbor supermassive black holes in their central
regions, and these black holes play a crucial role in the evolution of their host galaxies.  The
underlying mechanism is known as active galactic nuclei (AGN) feedback, which implies that the
changes of the density and temperature of  ISM in the galaxy due to the radiation and outflow from
the AGN and subsequently the changes of star formation and black hole fueling  \citep[and references
therein]{Fabian:12,Kormendy:13}.

To evaluate the effects of AGN feedback on galaxy evolution by numerical simulations, ideally
we should simulate from the central black hole to the whole galaxy, from the BH Schwarzschild
radius of $R_{s}\sim 10^{-5}\,{\rm pc} \, (M_{\rm BH}/10^{8}\,M_{\odot})$ to the galaxy scale
of $\sim 100s\,{\rm kpc}$.  However, the ratio of the size scale of the black hole and the host
galaxy is more than ten orders of magnitudes, thus  it is not feasible to cover such a large
dynamical range even with a state-of-art supercomputer.
Therefore, different works in the literature focus on different spatial scales. Most of the works focus on the
scale much larger than the black hole accretion flow. In this case,  the AGN is difficult
to resolve and a so-called ``sub-grid'' model to describe the AGN physics has been developed and
widely used in many simulation works.

It is obvious that in this case, it is crucial to have the correct description of the output
from the central AGN. Then the first important parameter is  the mass accretion rate of
the AGN. Many works focus on very large scales so that it is difficult to resolve the Bondi
radius and  calculate the accretion rate directly. In this cases, the accretion rates have to
be estimated in some way thus the value  can be underestimated or overestimated significantly
(e.g., see \citealt{Negri:17} and references therein, see also \citealt{Korol:16,Ciotti:17a})
Some other works, which focus on relatively smaller scales, overcome this problem by resolving the
Bondi radius \citep{Ciotti:97, Ciotti:01, Ciotti:07, Ciotti:09b, Shin:10, Ostriker:10, Novak:11,
Novak:12, Gan:14, Ciotti:17b}. In these works, the inner boundary is set to be a few pc, which
is smaller than the Bondi radius, so we can directly calculate (rather than estimate) the mass
accretion rate at the inner boundary.  The outer boundary is large enough to reach $\sim 100s$ kpc,
allowing the study of the evolution of the whole galaxy  and even the circum-galactic medium (CGM).

Once the mass accretion rate is reliably calculated, the output of the AGN, namely the radiation
and wind (jet is neglected in our work), is determined by the accretion physics adopted (e.g.,
see the review of accretion theory by \citealt{Yuan:14}). Most recently,  by taking into account
the recent developments of the theory of black hole accretion, \citet[hereafter Paper I]{Yuan:18}
have presented the most updated descriptions of the AGN outputs. In this work, the authors
have investigated the AGN feedback effects in an isolated elliptical galaxy by performing
two dimensional hydrodynamical numerical simulations. The inner boundary of the simulation
is chosen so that the Bondi radius is resolved and the accretion rate is precisely determined.
They discriminate between the cold and hot accretion modes according to the value of the accretion
rate, and present proper descriptions of wind and radiation emitted from the accretion flow
in the two modes. Their numerical results indicate that these updates of the AGN physics are
crucial for determining the effects of AGN feedback.  The updated AGN physics is described
in detail in \citetalias{Yuan:18} and will be briefly reviewed it in \S~\ref{sec:AGNphysics}
of the present paper.

In \citetalias{Yuan:18}, the specific angular momentum of the gas in the galaxy is assumed
to be very small. Although elliptical galaxies are primarily  pressure-supported rather than
rotation-supported, the angular momentum of the gas in many galaxies is likely not small. Recent
kinematic surveys have revealed that $\sim$80\% of early-type galaxies (ETGs) belong to regular
rotators, characterized by oblate axisymmetric shapes that reflect underlying disk-like components
\citep{Krajnovic:11, Krajnovic:13, Emsellem:11}.  Even in the slowly rotating ETGs, a mid-plane
HI disk, indicative of the presence of a stellar disk, is frequently detected \citep{Serra:14}.

The goal of this work is to extend \citetalias{Yuan:18} by considering the case of higher angular
momentum. In this case, the new physics required to include in the study compared to
\citetalias{Yuan:18} is the physical mechanism of angular momentum transport.  As we will describe
later, it seems as if different mechanisms play their respective roles on different scales. On the
galactic scale, one of the most promising mechanism is the gravitational torque raised by the
non-axisymmetric structure of stars in the galaxy, proposed in \citet{Hopkins:10,Hopkins:11}. Since
our simulation is two dimensional, it is impossible to include such a mechanism from first
principles.  Instead we adopt a phenomenological approach by using the $\alpha$-description, similar
to the case of black hole accretion disk \citep{Shakura:73}.  This $\alpha$-description was devised
with a physically motivated, dimensionless scaling of the kinematic viscosity such that the strength
of the angular momentum transport process would be represented in dimensionless fashion by $\alpha$.
The detailed implement in our work is presented in \S~\ref{subsec:hydro}.

In the present work, we ignore AGN activity triggered by an external mechanism, such as galactic
mergers. Although the mergers may be effective in triggering AGN activity  \citep{Mihos:96,
DiMatteo:05},  there are reasons that the assumption of an isolated galaxy is worth
considering. First, the observations indicate that BH growth in massive galaxies is likely  driven by internal
secular processes rather than by significant mergers, at least since redshift $z \sim 2$
\citep{Schawinski:11, Kocevski:12, Fan:14}.  Secondly, even in the absence of merging, the
total amount of gas injected from the pure stellar evolution is large enough to produce a BH
two orders of magnitudes more massive than what is observed \citep{Ciotti:12}.
In the future, we will examine the effect of the galactic merging on the evolution
of the central black hole and its host galaxy. It is speculated that the
gas contents  may be enhanced, which then possibly induce both star formation and AGN  activity especially in early
evolution time. 

As the second paper of a series of project, in this work we extend the work of \citetalias{Yuan:18}
to the case of an elliptical galaxy which is partly rotation-supported rather than fully
pressure-supported.  The paper is organized as follows. In \S~\ref{sec:galaxy}, we describe
the detailed model of the galaxy. In \S~\ref{sec:AGNphysics}, we briefly review the updated AGN
physics that we adopt in the simulation.  In \S~\ref{sec:setup}, we present the numerical
setup and treatment of angular momentum transport.  In \S~\ref{sec:results}, we describe the
simulation results, including  the AGN light curve, growth of black hole mass, star formation,
and X-ray emission of the galaxy.  Summary and conclusion are presented in \S~\ref{sec:summary}.

\begin{deluxetable}{ccccc}[!htbp]
    \tablecolumns{5}\tabletypesize{\scriptsize}\tablewidth{0pt}
    \tablecaption{Description of the simulations \label{tab:model}}
    \tablehead{ \colhead{model} & \colhead{k} & \colhead{$\alpha_{\rm visc}$} & \colhead{Mechanical} & \colhead{Radiative} \\
                \colhead{} & \colhead{} & \colhead{} & \colhead{Feedback} & \colhead{Feedback}}
    \startdata
        k00$^{1}$ & 0   & 0.1  & o & o \\
        k01       & 0.1 & 0.1  & o & o \\
        k03       & 0.3 & 0.1  & o & o \\
        k05$^{2}$ & 0.5 & 0.1  & o & o \\
        k07       & 0.7 & 0.1  & o & o \\
        k09       & 0.9 & 0.1  & o & o \\
        k05noFB   & 0.5 & 0.1  & x & x \\
        k05windFB & 0.5 & 0.1  & o & x \\
        k05radFB  & 0.5 & 0.1  & x & o \\
        k05alp-2  & 0.5 & 0.01 & o & o \\
    \enddata
    \tablenotetext{1}{k00 is the same model with fullFB in \citetalias{Yuan:18}.}
    \tablenotetext{2}{fiducial model in this paper.}
\end{deluxetable}

~~~~

\section{Galaxy Models}\label{sec:galaxy}

In this section, we briefly introduce the key features of the galaxy models, with respect to the
stellar population and evolution, the dynamical structure, and the galactic rotation.  To isolate
the problem, we set the many aspects of the simulations to be same as that in \citet[see also
\citealt{Novak:11} for more detailed description]{Gan:14}, except for the treatment of galactic
rotation (see \S\ref{subsec:galactic_str}).

~~~~

\subsection{Stellar Evolution}\label{subsec:star_evol}

In elliptical galaxies, the gas is supplied by evolved stars predominantly in the phases of the
red giant, asymptotic giant branch, and planetary nebula. The total mass of gas injected from
the pure passive stellar evolution is two orders of magnitudes larger than the BH mass observed
in elliptical galaxies \citep{Ciotti:12}. In our simulation, the gas density is initially set
to be low so that the gas is mainly supplied from stellar evolution, i.e. ``secular evolution''.

Following the description in \citet{Pellegrini:12}, the total mass loss rate of a stellar population is computed by
\begin{equation}\label{eq:Ms}
    \dot{M}(t) = \dot{M}_{\star}(t) + \dot{M}_{\rm SN}(t),
\end{equation}
where $\dot{M}_{\star}$ is the mass loss rate for an evolved star, which is approximated as the
single burst stellar population synthesis model \citep{Maraston:05},
\begin{equation}
    \dot{M}_{\star} = 10^{-12}\,A\times M_{\star}\,t_{12}^{-1.3} ~~ M_{\odot}\,\rm yr^{-1},
\end{equation}
where $M_{\star}$ is the galactic stellar mass in solar mass unit at an age of 12 Gyrs, $t_{12}$
is the age in units of 12 Gyrs. For all models, we set the galactic stellar mass to $M_{\star}
= 3\times 10^{11}\,M_{\odot}$ and the coefficient $A$ is set to be 3.3, which is indicative
of Kroupa initial mass function. The recycled rate of gas from SNIa is $\dot{M}_{\rm SN} =
1.4\,M_{\odot}\,R_{\rm SN}(t)$, where $R_{\rm SN}$ is the evolution of the explosion rate with
time. The approximate mass loss rate, eq.~(\ref{eq:Ms}), is reliable for solar metal abundance.


\subsection{Galactic structure}\label{subsec:galactic_str}

The galaxy models are built following the procedure described elsewhere \citep{Ciotti:09a}, and
refer to an isolated elliptical galaxy placed on the Fundamental Plane with a projected stellar
velocity dispersion, considering spherically symmetric dark matter halo and stellar profile.
The stellar density profile is described by the Jaffe profile \citep{Jaffe:83},
\begin{equation}
    \rho_{\star} = \frac{M_{\star}\,r_{\star}}{4\pi r^{2} (r_{\star}+r)^{2}},
\end{equation}
where $M_{\star}$ and $r_{\star}$ are the total stellar mass and the scale length of the
galaxy, respectively. In this paper, we set the total stellar mass to $M_{\star} = 3\times
10^{11}\, M_{\odot}$ and the scale length to $r_{\star} = 9.2$ kpc, which corresponds
to the projected half-mass radius (i.e., effective radius) of $r_{e}=0.7447\,r_{\star} =
6.9$ kpc.  The dark halo profile is set by the total density profile
scaling as $\rho \propto r^{-2}$ at large radii, which is consistent with observed profiles
\citep{Czoske:08,Dye:08,Auger:10,Sonnenfeld:13}.

We adopt the central velocity dispersion of $\sigma_{0} = 260\,\rm km\,s^{-1}$. For the minimum
halo model, the systematic rotational velocity, $v_{c}$ of the galaxy is
computed as
\begin{equation}
    v_{c} = \sqrt{2} \sigma_{0} = 368 \,\rm km\,s^{-1}.
\end{equation}
In this model, the gas flows under the total gravitational potential of
\begin{equation}
    \Phi_{\rm tot} = v_{c}^2 \,\ln \left( \frac{r}{r_{\star}} \right).
\end{equation}

As discussed in \S~\ref{subsec:star_evol}, most of gas is supplied by stellar evolution.
Therefore, the angular momentum of the gas, which is ejected through the stellar wind (i.e.,
mass losses from asymptotic giant branch stars), is initially set by the rotating motion
of stellar components. We introduce the rotation factor, $k$, to set the degree of the
stellar rotation, which is described as,
\begin{equation}
    v_{\phi,\star} = k\, \sin\theta\, \sigma_{r},
\end{equation}
where $\sigma_{r}$ the isotropic 1-dimensional stellar velocity dispersion without the contribution of
the central black hole,
\begin{equation}
    \sigma_{r}^{2} = \sigma_{0}^{2}(1+s)^{2}s^{2}
                    \left[ 6 \ln\left( \frac{1+s}{s} \right) + \frac{1-3s-6s^{2}}{s^{2}(1+s)} \right],
\end{equation}
where $s\equiv r/r_{\star}$. We perform the subsets of simulations with $k=$ 0.1, 0.3, 0.5, 0.7,
0.9 (see Table~\ref{tab:model}), and thus the supplied gas from the stellar evolution rotates
with sub-Keplerian velocity. However, as it falls into the mid-plane disk, the angular momentum
of the infalling gas reaches the Keplerian value (see Figure~\ref{fig:disk}).

From the perspective of energetics, the thermalization of the stellar mass losses due to the
stellar velocity dispersion should be taken into account \citep{Parriott:08}. The amount of
the thermalization heating decreases as the stellar rotation velocity increases since the
ordered motion becomes dominant, compared to the random motion of the stars. The detailed
description of the thermalization is discussed in \S~\ref{subsec:hydro}.

A central black hole also contributes to the gravitational potential, with its
dominance limited to the central region. The initial BH mass is chosen by the empirical correlation
between the black hole mass and the bulge mass in \citep{Kormendy:13}, which is
\begin{equation}
    M_{\rm BH} = 5\times 10^{8}\, M_{\odot} \, \left( \frac{M_{\star}}{10^{11}\,M_{\odot}} \right)^{1.17}.
\end{equation}
Given our fiducial galactic stellar mass, the initial black hole mass is set to $M_{\rm BH,init} =
1.8\times 10^{9}\,M_{\odot}$ for all models.\footnote{We note here that there is some uncertainty
in this value and eq.~(12) of \citet{Kormendy:13}: for our chosen value of $\sigma_{0}=260\,\rm
km\,s^{-1}$, the corresponding black hole mass is $M_{\rm BH}=9\times 10^{8}\,M_{\odot}$.}

Various shapes and internal kinematics of galaxy models were taken into account in the previous
study of X-ray haloes in ETGs \citep{Negri:14}. In that work, they adopted axisymmetric galaxy
models, for which the galaxies are flattened by either non-isotropic stellar velocity dispersion
or rotation. However, they ignored the effects of AGN feedback, which likely plays a significant
role in producing X-ray radiation in the central region \citepalias[see][]{Yuan:18}.  As a
following work, \citet{Ciotti:17b} applied this axisymmetric galaxy model to the numerical study
of AGN feedback. In that work, the flattening was set to occur by the non-isotropic stellar
velocity dispersion without consideration of the galactic rotation. Although our galaxy model
is spherically symmetric to isolate the problem with \citetalias{Yuan:18}, in the present work,
we consider both galactic rotation and AGN feedback to understand their respective roles in
the evolution of the black hole and its host galaxy.  In the following study, the axisymmetric
galaxy models will be also taken into account.

\subsection{Angular momentum transport}

The inflow of gas into the galactic center is essential to manipulate active star formation and
switch on the AGN activity. The triggers of such inflows are diverse and the dominant one depends
on the scale: on a galactic scale, tidal torques driven by major mergers or cosmological infall
lead to rapid gas inflow into the central $\sim$ kpc \citep{Hernquist:89, Barnes:91, Barnes:96}.
Minor mergers and/or disk instabilities, which cause bar and spiral structures, may also produce
similar gravitational torques \citep{Hernquist:95,Bournaud:05,Younger:08}.  Once gas reaches
sub-kpc scales, it cools rapidly, which can cause a non-axisymmetric gravitational instability
and torque, and thus a large fraction of the gas can flow toward the central region at $\sim$
0.01 pc \citep{Bertin:01,Lodato:04,Hopkins:10,Hopkins:11}. Near the central BH, it is believed
that MHD turbulence, produced by magnetorotational instability, is responsible for the angular
momentum transport \citep{Balbus:91,Balbus:98}.

In our simulation, we consider the secular evolution of elliptical galaxies without merger events.
As we will discuss, for most of our simulation, as the gas cools down, it forms a mid-plane
disk at the scale of $\sim$ kpc, within which the gravitational torques by non-secondary
instabilities are favored to be the dominant mechanism for the angular momentum transport
\citep{Hopkins:10,Hopkins:11}.  However, since we perform the simulations in two dimensions
without considering self-gravity of the gas, we technically cannot include the non-axisymmetric
torque. As an alternative approach, we make use of  stress tensor, $\boldsymbol{T}$, with the
viscosity $\alpha$-prescription \citep {Shakura:73} to mimic the gravitational disturbances
exerting on the gas.  (see \S~\ref{subsec:hydro} for the detailed description). While the proper
$\alpha_{\rm visc}$ value is largely uncertain, we set it to the fiducial value of $\alpha_{\rm
visc}=0.1$ for most runs.  For this value of $\alpha$, in our fiducial model, k05, the gas
inflows at $\sim$ 100s pc  of disk is $0.1\sim10 M_{\odot}\,\rm yr^{-1}$. This value is roughly
consistent with the median value of the inflow rate obtained in \citet{Hopkins:11},
although the latter has larger scatter.


\citet{Hopkins:10} showed that a gas-rich galaxy merger can induce the gas inflow rate up to
$100\, M_{\odot}\,\rm yr^{-1}$ at 300 pc. Compared to the case of gas-rich galaxy merger, it is
not surprising that gas inflow rate in our results is small because we focus on the ``secular''
evolution of galaxy, which is gas-poor initially. Even for the model without galactic rotation,
the model k00, the large-scale inflow is a few $10s\, M_{\odot}\,\rm yr^{-1}$, which is larger
than in the model k05, but is still smaller than in the case of gas-rich merger, mainly due to
the dearth of initial gas in our galactic model.  However, as we will discuss, in the present
work, the gas inflow rate is large enough to trigger a luminous AGN.

~~~~

\section{Physical model of AGN Feedback}\label{sec:AGNphysics}

We adopt the most updated ``sub-grid'' AGN physics presented in \citetalias{Yuan:18}.
For completeness, we briefly review it as follows.

The first important thing to note is that depending on the value of accretion rate, black hole
accretion is divided into two modes. When the accretion rate is relatively high, it belongs to the
cold accretion mode.  The most luminous AGN, such as quasars, are in this mode. When the accretion
rate is lower, it shifts to the hot accretion mode, in which AGNs spend most of their time.
The two modes are bounded by a critical luminosity,
\begin{equation}
    L_{\rm c}\approx 2\%\,L_{\rm Edd}.
\end{equation}

The cold and hot accretion modes are described by the standard thin disk \citep{Shakura:73}
and the hot accretion flows, respectively \citep{Yuan:14}. These are two completely different
accretion modes in the sense of the dynamics and radiation of the accretion flows. For the cold
mode, many mechanisms seem to play a role in producing wind, such as thermal, magnetic, and
radiation (line-force) \citep[see][]{Proga:03}, and theoretically the wind production is  still a
partially solved problem. Therefore, we adopt the observational results of \citet{Gofford:15} to
describe the wind properties. The mass, momentum and energy fluxes of the winds are described by,
\begin{equation}
    \dot{M}_{\rm W,C} = 0.28 \left( \frac{L_{\rm bol}}{10^{45}\,\rm erg\,s^{-1}} \right)^{0.85} \, M_{\odot}\,\rm yr^{-1},
\end{equation}
\begin{equation}
    \dot{P}_{\rm W,C} = \dot{M}_{\rm W,C}\,v_{\rm W,C},
\end{equation}
\begin{equation}
    \dot{E}_{\rm W,C} = \frac{1}{2}\dot{P}_{\rm W,C}\,v_{\rm W,C},
\end{equation}
where $L_{\rm bol}$ is the bolometric luminosity of the AGN. The velocity of wind is described by,
\begin{equation}
    v_{W,C} = 2.5\times10^{4}\, \left( \frac{L_{\rm bol}}{10^{45}\,\rm erg\,s^{-1}} \right)^{0.4}\, \rm km\,s^{-1},
\end{equation}
and we set the maximum wind velocity of $10^{5}\, \rm km\,s^{-1}$.

The radiation from the thin disk is well known, and is approximated as,
\begin{equation}
    L_{\rm bol} = \epsilon_{\rm EM,cold}\,\dot{M}_{\rm BH} c^{2},
\end{equation}
where $\dot{M}_{\rm BH}$ is the BH accretion rate, $c$ is the speed of light, and $\epsilon_{\rm
EM,cold}$ is the radiative efficiency. Here we assume the value of the efficiency in cold mode
is 0.1, implying that the BH is moderately spinning. This value also agrees with the empirical
studies of \citet{Yu:02} and \citet{Soltan:82}. In addition to luminosity, another important parameter
to describe the radiative heating to the ISM of the host galaxy by Compton scattering is the
Compton temperature of the radiation. In the cold mode, its value is $T_C\approx 10^7{\rm K}$
\citep{Sazonov:04}.

The black hole mass accretion rate in the cold mode is calculated by
\begin{equation}
    \dot{M}_{\rm BH,cold} = \dot{M}_{\rm d,inflow} - \dot{M}_{\rm W,C},
\end{equation}
where $\dot{M}_{\rm d,inflow}$ is the mass inflow rate in the accretion disk around the black hole with
the instantaneous viscous timescale, $\tau_{\rm visc}\approx 10^{6}$ yr \citepalias[see][for more detailed description]{Yuan:18}.

In the hot accretion mode, the geometry of the accretion flow is usually an inner hot accretion flow
plus an outer truncated thin disk \citep{Yuan:14}. The truncation radius is described by,
\begin{equation}\label{eq:truncR}
        R_{\rm tr} \approx 3\,R_{s} \left[ \frac{2\times10^{-2}\,\dot{M}_{\rm Edd}}{\dot{M}(r_{\rm Bondi})} \right]^{2},
\end{equation}
where $R_{s}$ is the Schwarzschild radius, which is $R_{s} \equiv 2\,G\,M_{\rm BH}/c^{2}$.
In contrast to the case of the cold accretion mode, wind production in the hot mode is theoretically
well studied \citep[e.g.,][]{Yuan:12, Yuan:15, Bu:16}  but the observational constraints
are much worse due to the lack of observational data. Using the trajectory approach, \citet{Yuan:15}
have carefully calculated the fluxes of mass, momentum, and energy of wind based on
the GRMHD simulation of black hole accretion:
\begin{equation}
    \dot{M}_{\rm W,H} \approx \dot{M}_{r_{\rm Bondi}} \left[ 1 - \left( \frac{3\,r_{s}}{r_{\rm tr}} \right)^{0.5} \right],
\end{equation}
\begin{equation}
    \dot{P}_{\rm W,H} = \dot{M}_{\rm W,H}\,v_{\rm W,H},
\end{equation}
\begin{equation}
    \dot{E}_{\rm W,H} = \frac{1}{2}\dot{M}_{\rm W,H}\,v_{\rm W,H}^{2},
\end{equation}
where $r_{\rm Bondi}$ is the Bondi radius and the wind velocity is approximated as
\begin{equation}
    v_{\rm W,H}\approx (0.2-0.4)\,v_{\rm K}(r_{tr}),
\end{equation}
where $v_{\rm K}$ is the Keplerian velocity.

The black hole accretion rate in the hot mode is computed by
\begin{equation}
   \dot{M}_{\rm BH,hot} \approx \dot{M}_{r_{\rm Bondi}} \left( \frac{3\,r_{s}}{r_{\rm tr}} \right)^{0.5}.
\end{equation}

Since hot accretion flows are optically thin, the radiation output from hot accretion flows
is much more complicated than that from the cold mode \citep{Yuan:14}. In this case,
the radiative efficiency is no longer a constant. The radiative efficiency
as a function of accretion rate is studied in \citet{Xie:12}, which gives the following
fitting formula
\begin{equation}
    \epsilon_{\rm EM,hot}(\dot{M}_{\rm BH}) = \epsilon_{0} \left( \frac{\dot{M}_{\rm BH}} {0.1\,L_{\rm Edd}/c^{2}}\right)^{a},
\end{equation}
where the value of $\epsilon_{0}$ and $a$ are given in \citet{Xie:12}. Here we summarize 
the set of ($\epsilon_0,\,a$) that is adopted for the current work.
\begin{eqnarray}
   (\epsilon_0, a) &=& \left\{ \begin{array}{ll} (0.2,0.59), & \dot{M}_{\rm BH}/\dot{M}_{\rm Edd}\la 9.4\times 10^{-5} \\
   (0.045,0.27), & 9.4\times 10^{-5} \lesssim \dot{M}_{\rm BH}/\dot{M}_{\rm Edd} \lesssim 5\times 10^{-3} \\
  (0.88,4.53), & 5\times 10^{-3}\lesssim \dot{M}_{\rm BH}/\dot{M}_{\rm Edd} \lesssim 6.6\times 10^{-3} \\
   (0.1,0), & 6.6\times 10^{-3}\lesssim \dot{M}_{\rm BH}/\dot{M}_{\rm Edd} \lesssim 2\times 10^{-2} \end{array} \right.
   \label{efficiencyfit}
\end{eqnarray}

The Compton temperature is higher than that of the cold mode, due to the difference of the emitted
spectrum between cold and hot modes. Its value is $T_c\approx 10^{8}$ K and $5\times10^{7}$ K for
the range of $10^{-3} \lesssim L/L_{\rm Edd} \lesssim 0.02$ and $L/L_{\rm Edd} \lesssim 10^{-3}$,
respectively \citep{Xie:17}.

~~~~

\section{Numerical Setup}\label{sec:setup}

We perform two-dimensional hydrodynamic simulations with ZEUS-MP2 \citep{Hayes:06} in spherical
coordinates (r,$\theta$). The grid resolution is 120x30.  The grid bin size in the radial direction
increases logarithmically and the range covers 2.5 pc $\sim$ 250 kpc.  A simulation with such
a large dynamical range is computationally expensive, hence we choose the two-dimensional simulation
assuming the axisymmetry. However, this approximation has a difficulty in resolving instabilities
and non-axisymmetric features such as spiral structures. In following work, we will extend this work
to three-dimensions.  We note that the Bondi radius is determined by the physical properties
at the galactic center: $r_{\rm Bondi} = G\,M_{\rm BH}/{c_{s,\rm in}^{2}} \approx 10\,{\rm pc} \,
(M_{\rm BH}/ 3\times10^{8}\,M_{\odot}) (T/ 10^{7}\,{\rm K})^{-1}$ \citep{Bondi:52}, where $c_{s,\rm
in}$ is the sound speed of the gas at the inner boundary. The Bondi radius varies during
simulations, and for most of time it is well resolved, which enables us to estimate the proper BH
accretion rate.


Galaxy mergers may be responsible for inducing star formation in the central region, and
fueling the BH and thus providing the power source to the quasar state at early cosmic times
\citep[e.g.][]{Mihos:96,DiMatteo:05,Cortijo:17}. The rapid growth of the BH and the formation
of elliptical galaxies possibly via merging process is beyond the scope of the current paper.
However, observations indicates that the hosts of AGNs are likely to evolve secularly rather than
being involved in an ongoing merger since z$\sim$2 \citep{Schawinski:11, Kocevski:12, Fan:14}.
We assume that in the simulation, the galaxy is well-established initially and evolves secularly.
The simulations begin at a galaxy age of $\sim$2 Gyr, which corresponds to the redshift $z\sim3$,
and for the comprehensive study, the evolution time spans 14 Gyr.

~~~~

\subsection{Hydrodynamics}\label{subsec:hydro}

The evolution of the galactic gas flow is computed by integrating the time-dependent Eulerian
equations for mass, momentum, energy conservations:
\begin{equation}
       \frac{\partial \rho}{\partial t} + \nabla \cdot \left( \rho \mathbf{v} \right) = \alpha_\star \rho_{\star} + \dot{\rho}_{II}-\dot{\rho}_{\star}^{+},
\end{equation}
\begin{equation}\label{eq:mom}
    \frac{\partial \mathbf{m}}{\partial t} + \nabla \cdot \left( \mathbf{m v} \right) = -\nabla p_{\rm gas} + \rho\mathbf{g} -\nabla p_{\rm rad} - \dot{\mathbf{m}}_{\star}^{+} + \nabla \cdot \boldsymbol{T},
\end{equation}
\begin{align}\label{eq:energy}
    \frac{\partial E}{\partial t} + \nabla \cdot \left( E \mathbf{v} \right) =& -p_{\rm gas}\nabla \cdot \mathbf{v} + H - C + \dot{E}_{I} + \dot{E}_{II}   \nonumber \\
              & + \dot{E}_{S} - \dot{E}_{\star}^{+}  + \boldsymbol{T}^{2}/\mu,
\end{align}
where $\rho,\,\mathbf{m},$ and $E$ are the gas mass, momentum and internal energy per unit volume,
respectively. The gas pressure is $p_{\rm gas} = \left( \gamma - 1 \right)\,E$, where the specific
heats is $\gamma = 5/3$. Here $\alpha_{\star} \rho_{\star}$ is the mass source from the stellar
evolution, and $\dot{\rho}_{II}$ is the recycled gas from supernovae (SNe) II.  We let the source
term $\alpha_{\star} \rho_{\star}$ evolve passively with Salpeter initial mass function
\citep{Salpeter:55}. $\dot{E}_{I}$ and $\dot{E}_{II}$ are feedback from SNe I and SNe II,
respectively: $\dot{E}_{I} = \dot{\rho}_{\rm Ia} \vartheta_{\rm SNIa}\,E_{\rm SN} /
(1.4\,M_{\odot})$, where the kinetic energy of a single SNIa is $E_{\rm SN} = 10^{51}\, \rm erg$ and
the thermalization efficiency $\vartheta_{\rm SNIa}=0.85$, of which we adopt the value as a
plausible one for low density and hot medium we adopt the value of 0.85
\citep[see][]{Mathews:89,Tang:05}.  When gas turns into stars, we eliminate the corresponding mass,
momentum, and energy ($\dot{\rho}_{\star}^{+},\, \dot{\mathbf{m}}_{\star}^{+},\,
\dot{E}_{\star}^{+}$) to conserve quantities, but also add new mass and energy from SNII explosions
($\dot{\rho}_{II},\, \dot{E}_{II}$). Under the assumption of Salpeter initial mass function, the
mass and the energy returned in SNII events for each star formation episode is 20\% of the newly
born star and their ejection timescale is $\tau_{II}=2\times 10^{7}$ years. All parameters in the
description of the stellar feedback are same for every models (see \citealt{Negri:15},
\citealt{Ciotti:17b} for more detailed description of stellar feedback).


In the energetics of the gas flows, the thermalization of the stellar mass loss, interacting
with the pre-existing hot ISM due to the stellar velocity dispersion, is important. Since in
our galactic model, the stars have a certain degree of the ordered motion, the thermalization
heating, $\dot{E}_{S}$, is computed by the trace of the velocity dispersion, which is expressed as,
\begin{eqnarray}
    \dot{E}_{S} &=& \frac{\alpha_{\star} \rho_{\star}}{2} \, \left[ {\rm Tr(\sigma^{2})} - v_{\phi,\star}^{2} \right]  \nonumber \\
                &=& \frac{\alpha_{\star} \rho_{\star}}{2}\, \left( 1 - \frac{\sin^{2}\theta\,k^{2}}{3} \right) \,{\rm Tr(\sigma^{2})},
\end{eqnarray}
in which the reduction factor $1-\sin^{2}\theta\,k^{2}/3$ takes into account the effect
of ordered rotation.

In eq.~(\ref{eq:mom}), we assume that the divergence of the tensor is strongest at the azimuthal
components, which can be approximated as
\begin{equation}\label{eq:partT}
    \nabla \cdot \boldsymbol{T} \approx \left( \frac{\partial T_{r \phi}}{\partial r} + \frac{1}{r}\frac{\partial T_{\theta\phi}}{\partial \theta}
        + \frac{3\,T_{r\phi}+2\,\cot{\theta}\,T_{\theta\phi}}{r}  \right)\,\hat{\phi},
\end{equation}
where, in spherical coordinates, the stress components are
\begin{equation}\label{eq:Trp}
    \boldsymbol{T}_{r\phi} \approx
        \mu \, r \frac{\partial}{\partial r} \left( \frac{v_{\phi}}{r} \right),
\end{equation}
\begin{equation}\label{eq:Ttp}
    \boldsymbol{T}_{\theta\phi} \approx
        \frac{\mu \, \sin{\theta}}{r} \frac{\partial}{\partial \theta} \left( \frac{v_{\phi}}{\sin{\theta}} \right).
\end{equation}
Here $\mu \equiv \rho\,\nu$ is the coefficient of shear viscosity, where $\nu$ is the kinematic viscosity coefficient,
\begin{equation}
    \nu = \alpha_{\rm visc}\,\frac{c_{s}^{2}}{\Omega_{\rm k}},
\end{equation}
where $c_{s}$ is the sound speed, $\Omega_{\rm k}$ is the Keplerian angular velocity,
and $\alpha_{\rm visc}$ is the viscosity parameter formulated by \citet{Shakura:73}.  Note that since
in this work, the simulation is performed with axisymmetric two dimensional coordinates,
we neglect the derivative terms of $\phi$ directions and approximate the tensors to
eqs.~(\ref{eq:Trp})\&(\ref{eq:Ttp}). The numerical treatment of angular momentum transport by
anomalous tensor is referred to \citet{Stone:99}.
We adopt the value $\alpha_{\rm visc}=0.1$ as a fiducial model, and for the comparison, we also perform
a subset of test simulations with an order of magnitude smaller value of $\alpha_{\rm visc}$ (see
\S~\ref{subsec:light}).

~~~~

\section{Results}\label{sec:results}

In this work, we allow for the possibility of high angular momentum of gas, deposited by the
rotating stars. The rotation speed of stars is determined by the parameter $k$, which varies from
0 (k00; no-rotation) to 0.9. Similarly with \citetalias{Yuan:18}, we also carry out a subset
of simulations to compare the results from one with both radiative feedback and mechanical
feedback (k00, k01, k03, k05, k07, k09), and one with purely radiative feedback (k05radFB),
and one with purely mechanical feedback (k05windFB), and one without AGN feedback (k05noFB).
Note that for the models with no/partial feedback, the rest configurations are same as k05. All
these models are listed in Table~\ref{tab:model}.

~~~~~

\subsection{Overview of the Evolution}\label{subsec:overview}

\begin{figure*}[!htbp]
    \centering
    \includegraphics[width=0.9\textwidth]{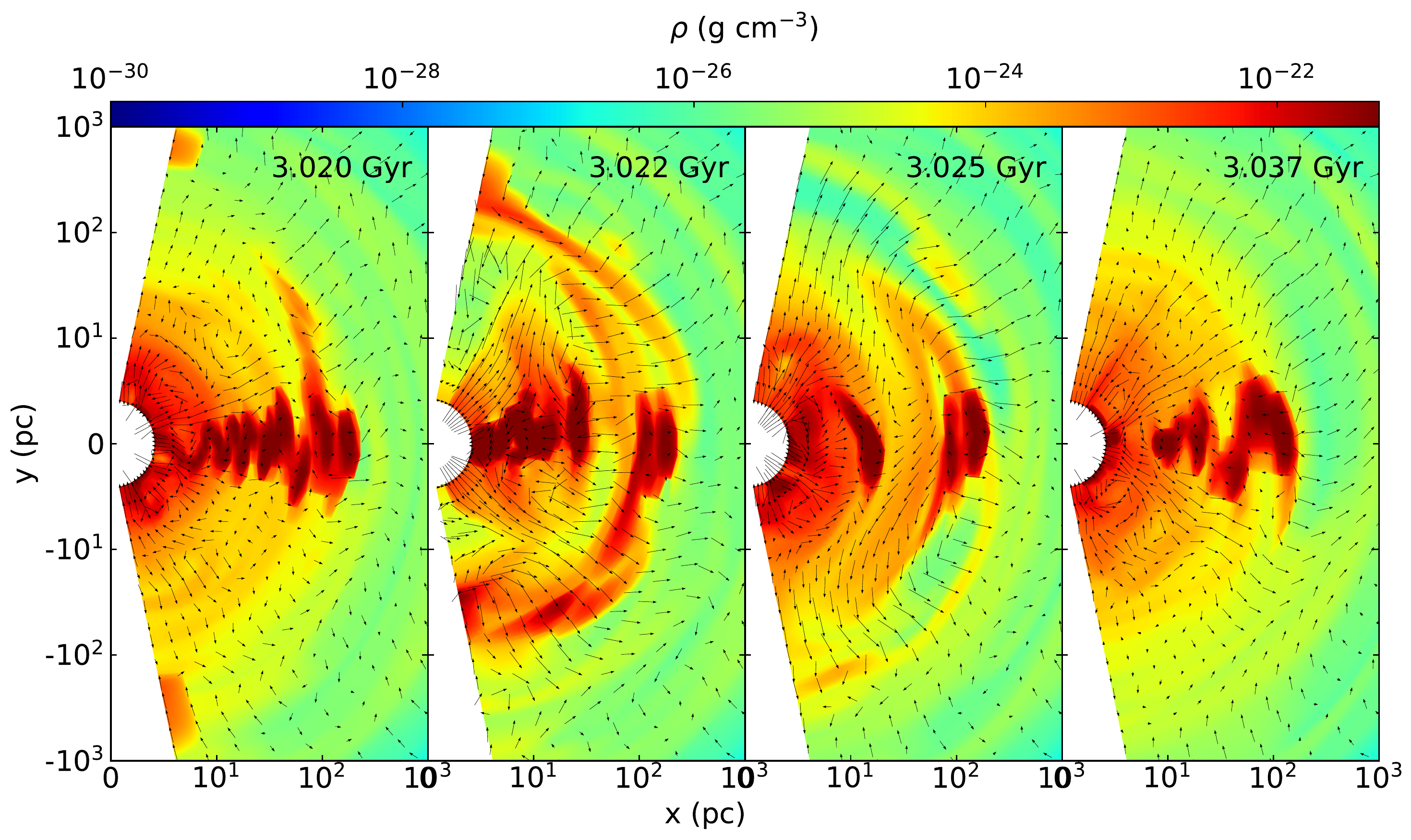}
    \caption{The time sequence of density map in one cycle of AGN burst for the fiducial model
            (k05).  The arrows in each plot indicate the velocity vectors. From the left, the
            density map corresponds to pre-burst (3.02 Gyr), ongoing-burst (3.022 Gyr), post-burst
            (3.025 Gyr), and quiescent (3.037 Gyr) periods, respectively. The length of velocity
            vectors is scaled with logarithmic in order to show the stream clearly.  \\~~~~}
    \label{fig:evol_c5}
\end{figure*}


In \citetalias{Yuan:18},  where the specific angular momentum of the gas in the galaxy is
assumed to be very small, the overall evolution of the AGN is as follows.  In the early stage, the
gas is enriched by stellar mass loss, and when it reaches sufficient density, the ISM undergoes
radiative cooling and produces cold shells and filamentary structures.  These structures
are unstable due to the Rayleigh-Taylor instability, then become disrupted in a short time.
The disrupted gas flows inward toward the center and triggers the AGN activity.   The strong
radiation and winds produced by the AGN heat the medium and expel the gas out of the central
region, and thus suppress the black hole accretion. The AGN activity is then reduced. As the ISM
is replenished by stellar winds and gradually cools down, the AGN cycle will start over. This is
the evolutionary track of the AGN cycle, which was inferred by the numerical results with the slowly
rotating (or non-rotating) galaxy models in the previous studies.

In the present paper, we focus on the case of much higher angular momentum of the gas in the
galaxy. While the general evolution picture of AGN activity is qualitatively similar, the details
are significantly different.   Figure~\ref{fig:evol_c5} shows the overview of AGN cycle in the
rotating galaxy. From the left plot, it shows the density map and gas streams in pre-burst,
ongoing-burst, post-burst, and quiescent periods, respectively.  While the gas with low angular
momentum accretes onto the BH from random directions, the gas with high angular momentum collapses
into the mid-plane and flows toward the center along the disk (left-most panel). Since the
density of the disk is high, on the path toward the BH, the gas is largely consumed by the
active star formation in the disk. The relative predominance between the BH accretion rate and
the star formation rate will be discussed in \S~\ref{subsec:starform}.

The gas in the disk likely loses its angular momentum via the various physical mechanisms, invoked
by multiple mechanisms such as magnetorotational instability \citep{Stone:01}, thermal
instability \citep{Bertin:01}, gravitational instability \citep{Toomre:64,Gammie:01} and non-axisymmetric
gravitational torque \citep{Hopkins:10,Hopkins:11,Angles-Alcazar:17}.  It is believed that no
single mechanism is dominant over the range of a few kpc to the BH event horizon, and
their relative importance depends on the scale of interest \citep[references
there in]{Dorodnitsyn:16}.
As the gas loses its angular momentum and accretes onto the central BH, the AGN activity
is triggered (second panel). Similar to the case of the non-rotating model, the AGN feedback
influences the surrounding medium by both the radiation and the mechanical winds. The radiation
heats the central region, within which the cold disk is likely photo-ionized and heated. The
AGN winds blowing out with a certain inclination angle is capable of generating turbulence,
which disturbs the cold disk inside-out (third panel). Recently, \citet{Tacchella:15} found
evidence that the star formation in the galactic disk is suppressed from the inside out. Our
numerical results indicate that the energy released from the central AGN drives such inside-out
star formation features. Finally, stellar mass loss replenishes the galaxy and the gas falls to
the mid-plane again (right-most panel), and a new AGN cycle starts over. We note that the key
difference of the AGN cycle between the slowly rotating galaxy and the rapidly rotating galaxy is
the presence of the mid-plane disk, which alters the gas fueling channel to the central black hole.
Many elliptical galaxies are indeed observed to contain central disks \citep[e.g.,][]{Serra:14}.

~~~~

\subsection{Light Curve of AGN Luminosity}\label{subsec:light}

\begin{figure*}[!htbp]
    \begin{center}$
        \begin{array}{cc}
            \includegraphics[width=0.5\textwidth]{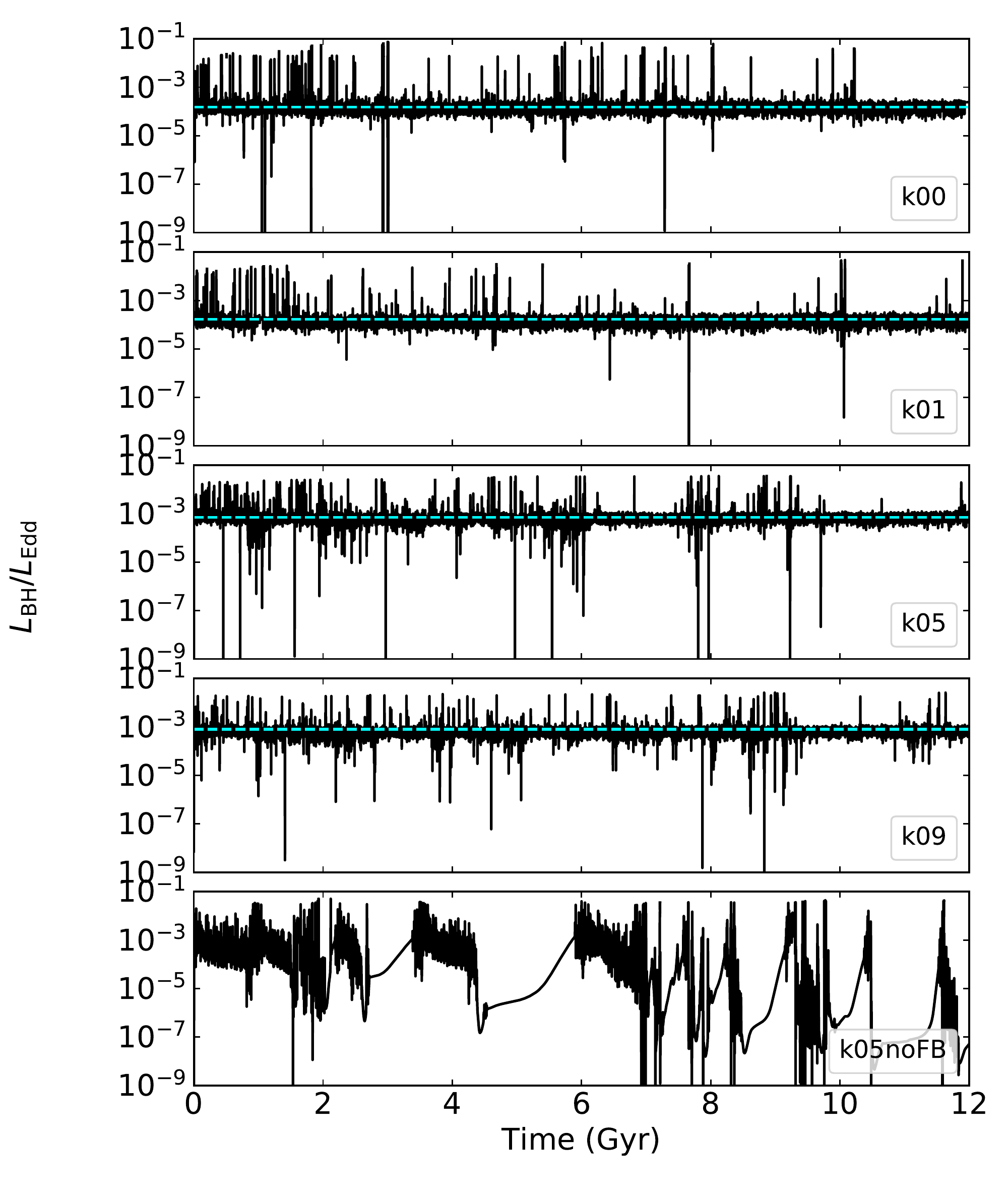} &
            \includegraphics[width=0.5\textwidth]{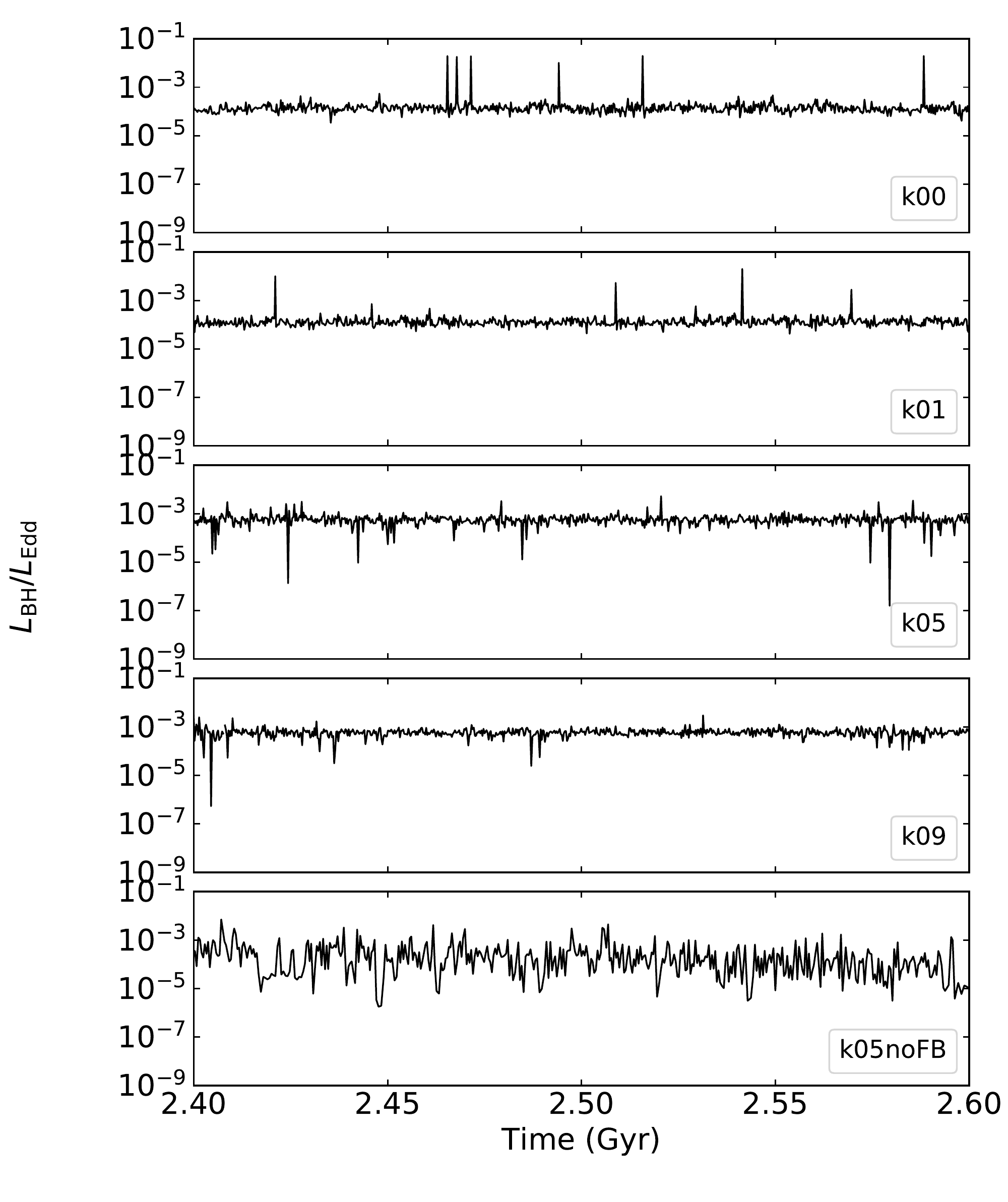}
        \end{array}$
    \end{center}
    \caption{Light curves of AGN luminosity as a function of time for the models with different
             amounts of galactic rotation.  The left panel shows the results in the entire
             evolution and the right panel shows the results over a shorter time interval. In
             the left panel, the horizontal cyan line represents the median value of the light
             curves.  \\~~~~}
    \label{fig:ldot}
\end{figure*}

Figure~\ref{fig:ldot} shows the evolution of light curves of AGN luminosity. The k00 model
represents the model without galactic rotation and k01, k05, k09 models represent the models,
in which the degree of the angular momentum increases in order.

The overall shape of the light curves are similar for every model with the full AGN feedback, indicating
that the effects of rotation on the galactic evolution may not be significant. However,
there are several aspects that show the trends, as the level of angular momentum is varied.
It is notable that as the galaxy rotates slower, the AGN burst occurs more frequently at early
evolution times, and the peak of the burst tends to be higher. This is because the lower angular
momentum provides a weaker barrier to hamper the BH accretion, producing the stronger AGN burst.

For most of the time, the AGN stays in the hot mode (i.e. low accretion regime; $L_{\rm BH} <
0.02\,L_{\rm Edd}$) as is observed.  In the left panel of Figure~\ref{fig:ldot}, the horizontal dashed lines
represent the mean value of the light curves in the hot mode. We found that the mean value
increases gradually as the host galaxy rotates faster. In the most rapidly rotating model, k09,
the mean value is a factor of 6 larger than in the non-rotating model, k00. This is because, in
the case of lower angular momentum, the AGN activity can reach to a higher level due to the
easier accretion of the gas. Consequently, the AGN can produce stronger radiation and winds,
which expel the gas surrounding the black hole more strongly. This results in a longer period
of time during which the accretion rate is very small.

In the right panel of Figure~\ref{fig:ldot}, we plot the AGN light curve over a shorter time
interval to compare the shape of the curve lines in different models. As discussed, all models
with AGN feedback have a similar shape, which is characterized by intermittent AGN bursts.
The typical life time of the AGN feedback is similar between the rotating models and the
non-rotating model, which is $\sim 10^{5}\,\rm yr$ (see \citetalias{Yuan:18} for the detailed
discussion).

In the model without AGN feedback, k05noFB, the light curve variability does not imply AGN
activity but simply indicates the variation of corresponding mass accretion rate.  While the mass
accretion rate for the model without feedback in a non-rotating galaxy shows monotonic decrease
(see noFB in \citetalias{Yuan:18}), the rate in model k05noFB still strongly fluctuates. This
is likely because, as in the case of black hole accretion flows, when the angular momentum of
the gas is present, the motion of the gas is convectively unstable, and such an instability
produces turbulence and episodic accretion.

In the model k05noFB, The light curve shows intermittent breaks (e.g. from 4.3 Gyr to 6 Gyr). The
reason is as follows. The density in the mid-plane disk is very high.  Without the influence of
AGN feedback, which can heat and disturb the disk by wind and radiation, the star formation and
thus the stellar feedback by the supernova are prone to be violent. Such violent stellar feedback
can destroy the disk and expel large amount of gas outward, which causes a strong decrease of
the black hole accretion rate. If the galaxy rotates slower, the disk will shrink and thus
the stellar feedback will be weaker. Therefore, these intermittent breaks take place only for
the rapidly rotating galaxy without AGN feedback. However, we may need to be careful that in
our two dimensional simulation, the stellar feedback on the disk can be overestimated. In the
case of three dimensional hydrodynamical simulations, the disk can fragment and thus the effects
of the stellar feedback may not be strong enough to expel the gas, forming galactic fountain
\citep{Biernacki:17}. We will investigate this issue in a following study having full three
dimensional simulation.

\begin{figure}[!htbp]
    \centering
    \includegraphics[width=0.5\textwidth]{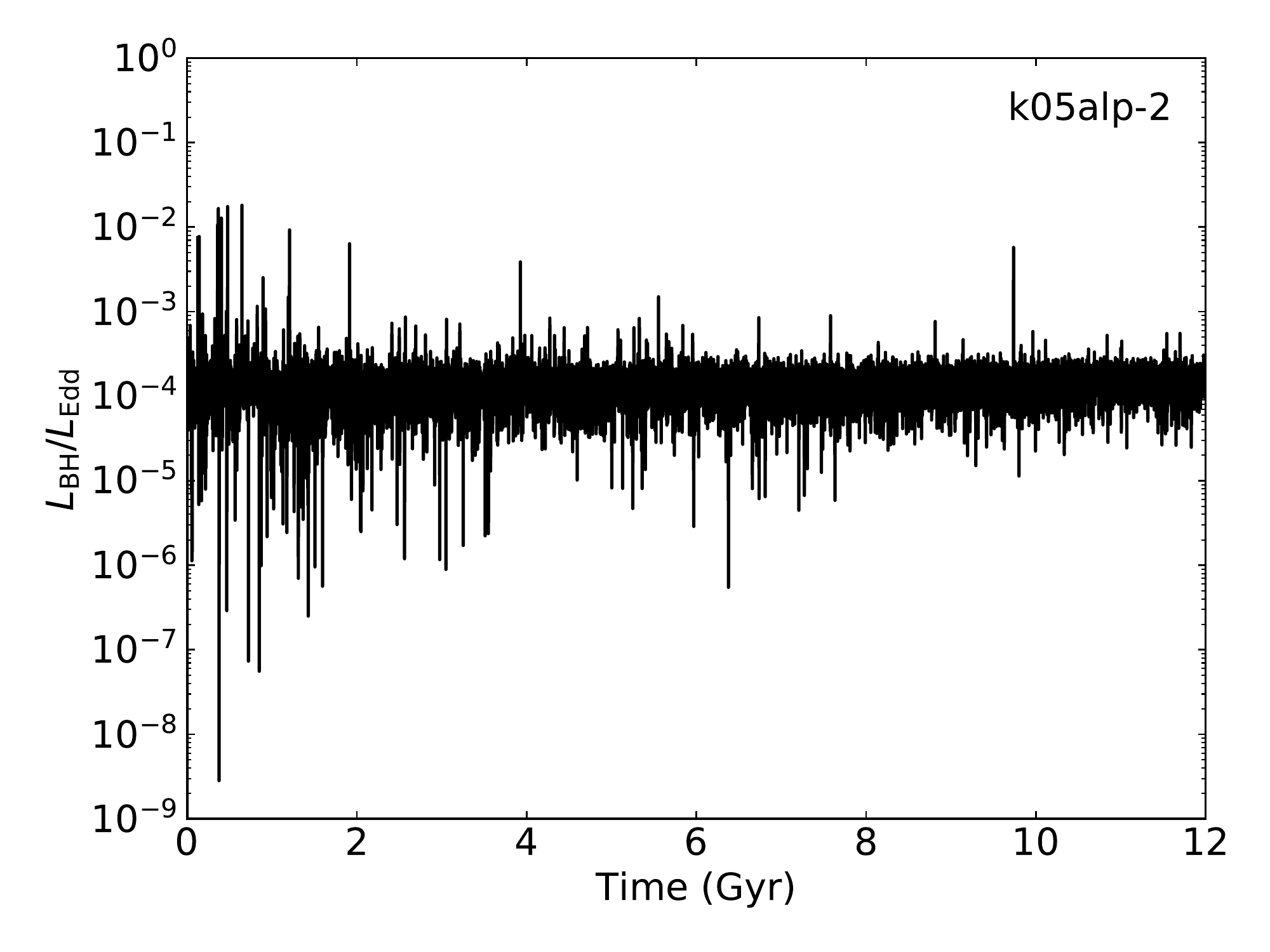}
    \caption{Light curve of AGN luminosity for the model with low viscosity $\alpha_{\rm visc}$
       parameter, $\alpha_{\rm visc}=0.01$  (model k05alp-2). \\~~~~ }
    \label{fig:ldot_alp2}
\end{figure}

The BH accretion and the feedback are manipulated by the inflow of gas. For the rotating galaxy,
such inflow is feasible only when the angular momentum of gas is transported outward. As a
result, the value of the viscosity parameter is of particular importance.  While the $\alpha_{\rm visc}$
value remains uncertain, we set the $\alpha_{\rm visc}$ value to 0.1 for most simulations.  For comparison,
we also carry out one run with $\alpha_{\rm visc}=0.01$ (k05alp-2) otherwise the same as the model k05.
Figure~\ref{fig:ldot_alp2} shows the AGN light curve for this model. Compared to the result
shown in Figure~\ref{fig:ldot}, we can see that the mean luminosity is significantly lower. This
is because, when $\alpha_{\rm visc}$ is smaller, the accretion timescale becomes longer, thus the gas
will stay in the disk for a longer time. Consequently, more gas will be consumed due to star
formation and thus the black hole accretion rate becomes smaller.

~~~~~

\subsection{Mass Growth of Black Hole}

\begin{figure}[!htbp]
    \centering
    \includegraphics[width=0.5\textwidth]{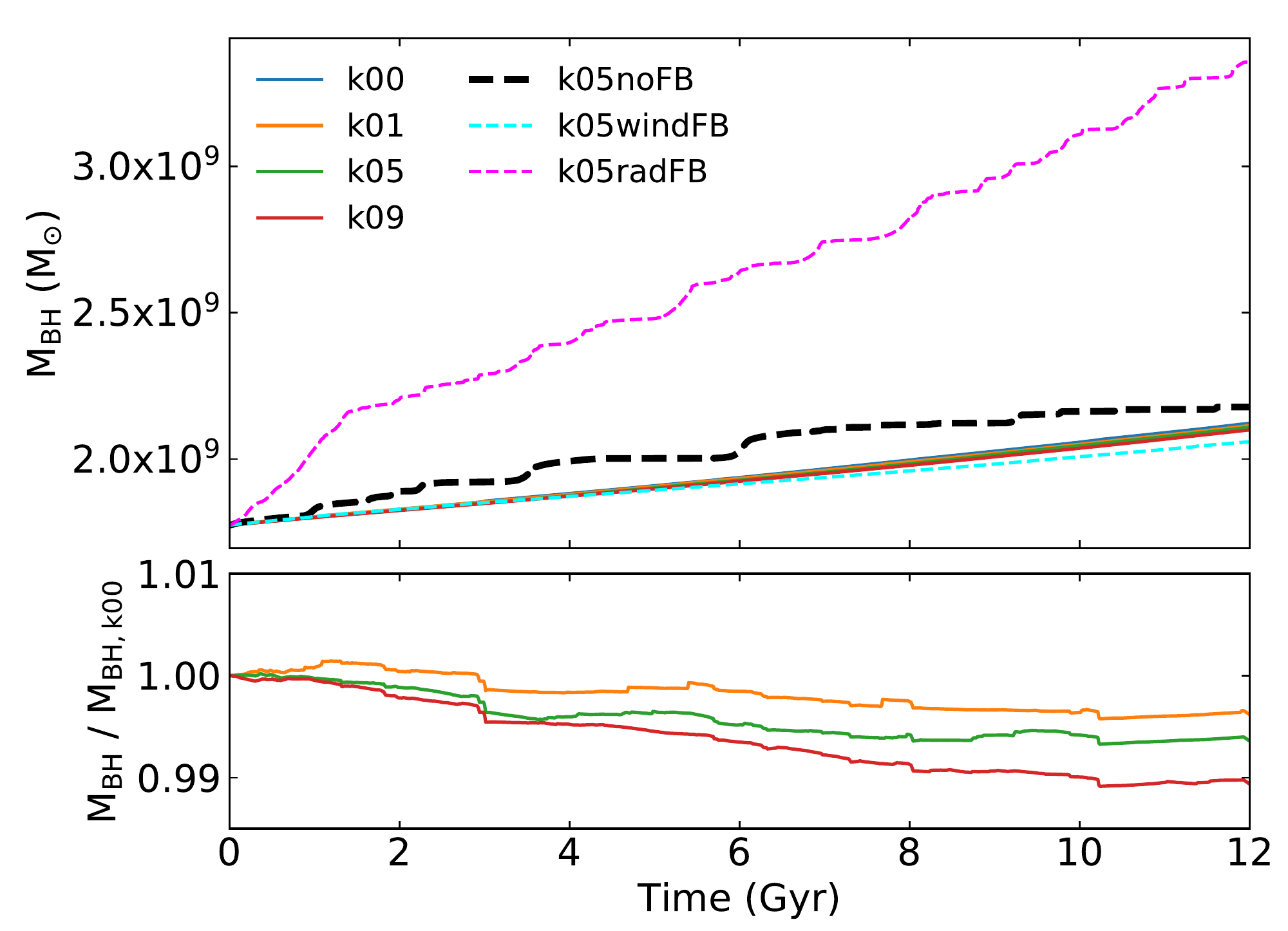}
    \caption{Mass evolution of BH for various models: solid lines represent the model with different galactic rotation.
            Black, cyan, and purple dashed lines indicate the BH mass growth for the model with no-AGN feedback,
            purely mechanical feedback, and purely radiative feedback, respectively. In bottom panel, the BH mass growth
            for k01, k05, k09 are normalized by $M_{\rm BH,k00}$. \\~~~~~}
    \label{fig:mbh}
\end{figure}

Mass accretion onto the BH is likely controlled by both AGN activity and galactic properties.
As discussed in \citetalias{Yuan:18}, the BH mass growth is regulated dominantly by mechanical
feedback (i.e. AGN wind), which expels gas out of central region during the bursts.  Unlike the
AGN wind, the irradiation by AGN plays a complicated role. The radiative pressure drives gas
outward, reducing the BH accretion. However, the radiative heating also suppresses the star
formation, preventing the gas depletion before it reaches the central BH, and thus increasing
the BH accretion simultaneously. The numerical results show that the latter effect is dominant,
hence the radiative feedback, in general, increases the BH growth rates.  In Figure~\ref{fig:mbh},
the final BH mass in the model with pure radiative feedback (k05radFB; of the magenta dashed line)
is two times  larger than that in the rest of models, implying that pure radiative feedback is
not effective in controlling the growth of the black hole mass. On the contrary, the mechanical
feedback plays the dominant role as we see that the BH mass growth in the model k05windFB is
similar to the full AGN feedback model (k05). This result was also found in the \citet{Choi:12}
simulations which had lower resolution but were 3D and included cosmological effects.  In our work,
we found that such primary roles for mechanical feedback and radiative feedback in regulating
the BH mass growth are similar for the models of non-rotating galaxy and rotating galaxy.

For the full feedback models, the difference of BH mass growth in the different galactic rotation
models may not be significant, however we found a monotonic change with the rotation speed. The
bottom panel of Figure~\ref{fig:mbh} shows the relative growth rate between the rotating galaxy
(k01, k05, k09) and the non-rotating galaxy (k00).  It is clear that as the galaxy rotates
faster, the BH accretion is more suppressed.  This is attributed to the angular momentum,
which prevents the gas from flowing inward.

\begin{figure}[!htbp]
    \centering
    \includegraphics[width=0.5\textwidth]{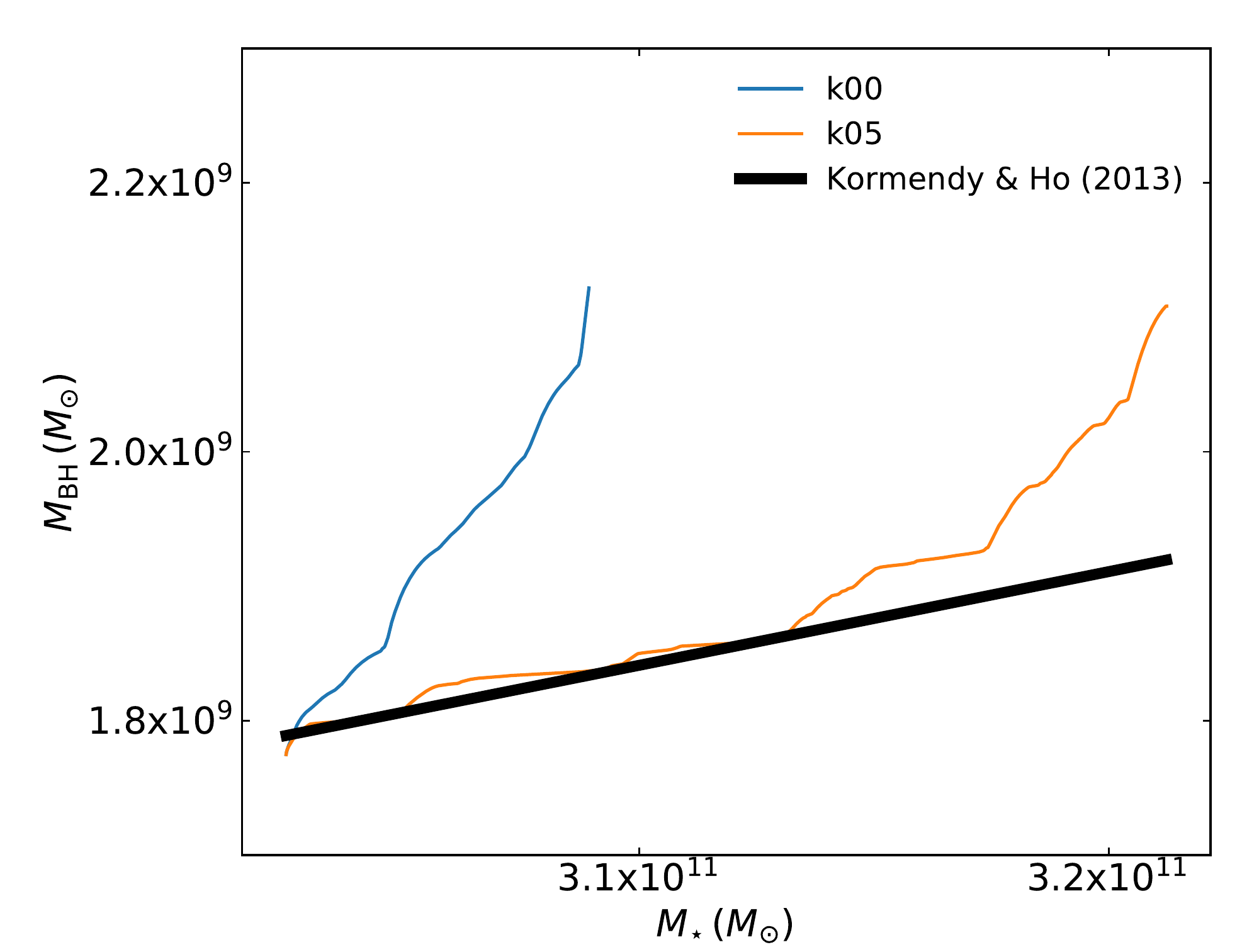}
    \caption{The relations between $M_{\rm BH}$ and $M_{\star}$. The blue and orange curves
            represent the results from the models k00 and k05, respectively. The thick black solid
            line represents the observed correlation of BH mass with bulge (elliptical) mass,
            which is derived with omission of pseudobulges \citep{Kormendy:13}.\\~~~~~}
    \label{fig:mbh_mst}
\end{figure}

It is believed that both the BH mass and the stellar mass may increase dramatically mainly through
galaxy merging at high redshift epochs \citep[e.g.,][]{Mihos:96,DiMatteo:05,Cortijo:17,Goulding:17}.
However, the mergers are unlikely to dominate BH growth and the $M_{\rm BH}-M_{\star}$ relation has
weak redshift evolution since $z \approx 2$ \citep{Kocevski:12,Fan:14,Yang:17}. Thus far, the
previous numerical studies for AGN feedback in an early type galaxy have a critical drawback that in
most of the results, the BH growth over the galactic evolution is considerably larger than what is
expected by the observed relation \citep{Kormendy:13}. For example, in \citet{Gan:14}, the final BH
mass of models with AGN feedback is $M_{\rm BH,final}~10^{9}-10^{10}\,M_{\odot}$, which is 5$\sim$30
times larger than the initial mass, while the stellar mass increases only several per cents of the
initial mass. In Figure~\ref{fig:mbh_mst}, we show the $M_{\rm BH}-M_{\star}$ relation from our
numerical data and the fitted formula from observation \citep{Kormendy:13}. We found that in our
updated model, the AGN feedback is effective in suppressing the BH growth and the result is
consistent with the expected relationship: in previous work, the ratio of BH mass growth, $\Delta
M_{\rm BH}/M_{\rm BH,init}$, is an order of magnitude larger than in the current result
\citep[e.g.,][]{Gan:14,Ciotti:17b}. In addition, we note that, when the host galaxy rotates faster,
the black hole mass growth is more suppressed as a consequence of more active star formation in the
mid plane disk, consuming more fuel before it accretes. (see \S~\ref{subsec:starform} for more
detailed discussion of the correlation between the angular momentum of the accreting gas and the
star formation). We argue that the scatters shown in the observed correlation between the black hole
mass and the stellar mass in the host galaxy may be ascribed to the degree of the galactic rotation.
This also may provide clues for the long-standing question {\it why and how BHs are no longer able
to grow above the critical value, $M_{\rm BH,max}\sim 10^{11}\,M_{\odot}$}, which is inconclusive
\citep[e.g,][]{King:16}. In order to shed light on this problem, larger parameter studies (e.g.,
various initial gas density) are required. We will discuss it in future work.


~~~~

\subsection{Star Formation}\label{subsec:starform}

\begin{figure*}[!htbp]
    \centering
    \includegraphics[width=0.9\textwidth]{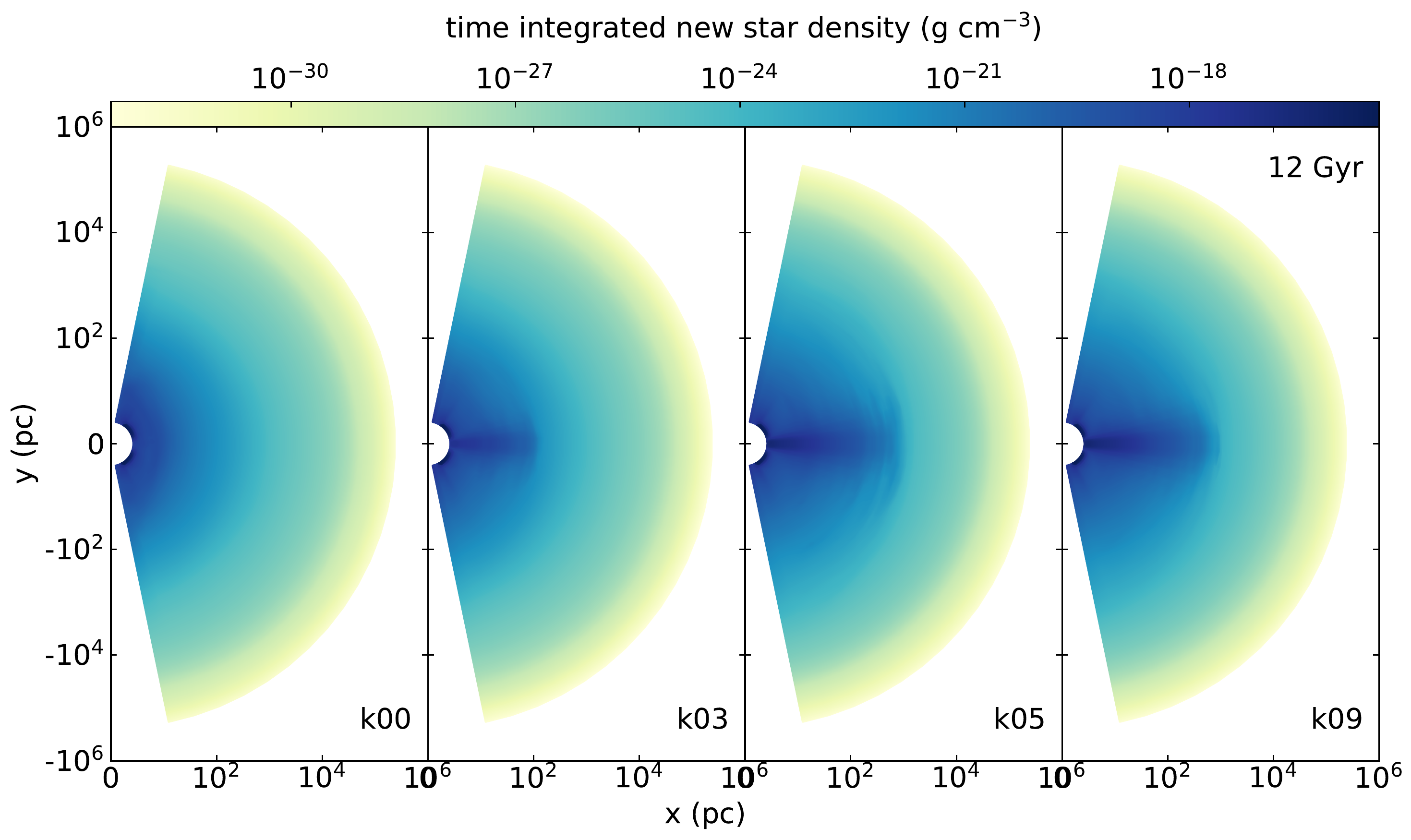}
    \caption{Time-integrated density of newly born stars at the end of the run for the models with different galactic rotation. \\~~~~}
    \label{fig:dnewst}
\end{figure*}

Figure~\ref{fig:dnewst} shows the density of the newly born stars, which is time-integrated up
to the end of runs. In a non-rotating galaxy (k00), stars form massively when the cold shells
and filaments fall back onto the central region, resulting in quite spherically symmetric and
centrally concentrated distribution (left-most column). However, as we can see the right three
columns, if the host galaxy rotates, stars form dominantly at the cold mid-plane disk, whose
size increases with the rotation speed.  The mid-plane disk spreads up to 0.1, 0.3, 0.5 kpc for
the models k03, k05, k09, respectively.

\begin{figure*}[!htbp]
    \begin{center}$
        \begin{array}{cc}
            \includegraphics[width=0.5\textwidth]{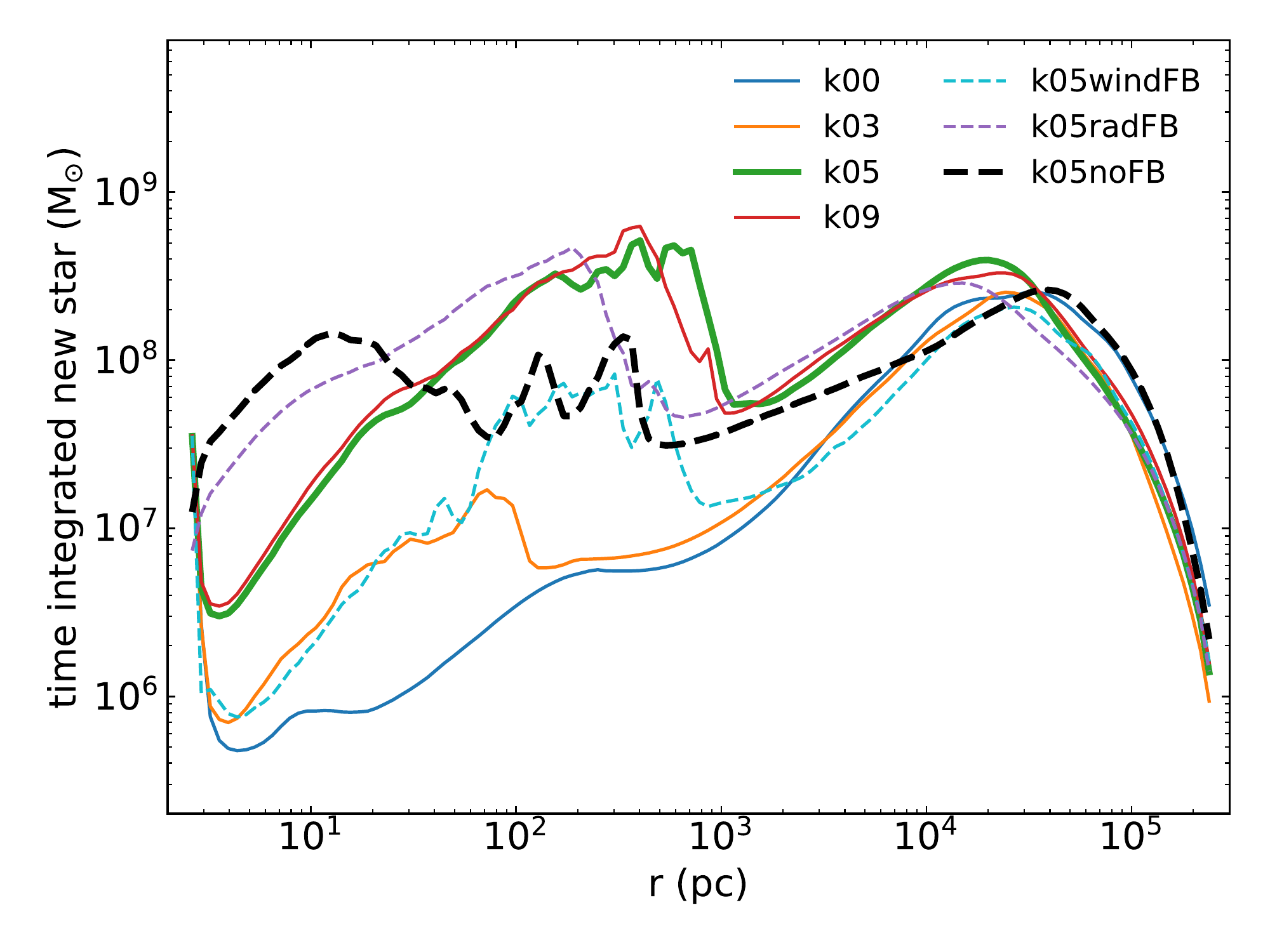} &
            \includegraphics[width=0.5\textwidth]{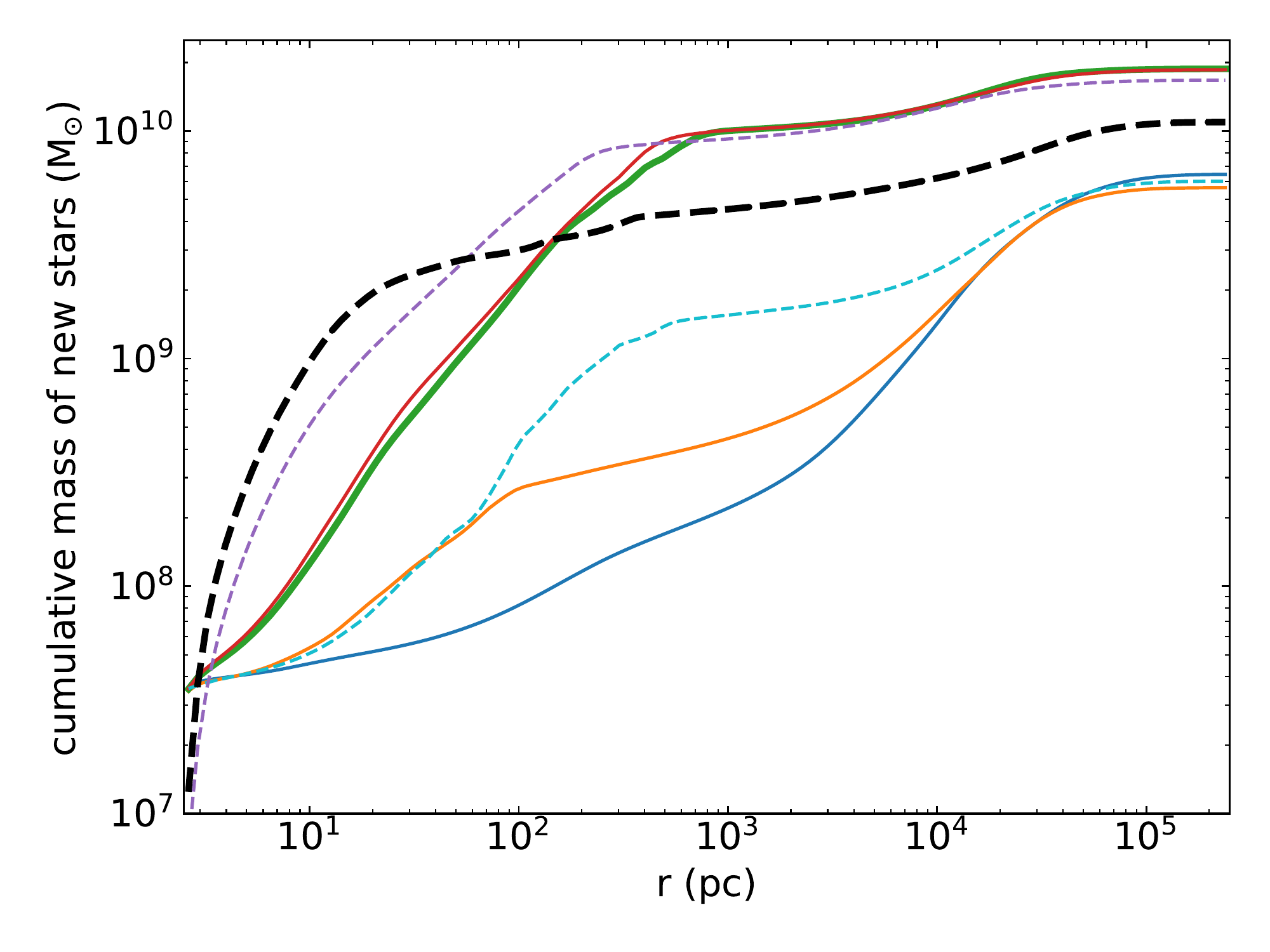}
        \end{array}$
    \end{center}
    \caption{{\it Left panel:} $\theta$-integrated mass of the newly born stars within each
             grid bin at a given radius at the end of the run.  {\it Right panel:} Enclosed
             mass of the newly born stars within a given radius at the end of the run. The two
             panels share the same denotation of lines. \\~~~~~}
    \label{fig:newst}
\end{figure*}

Figure~\ref{fig:newst} shows the time-integrated mass of newly born stars as a function
of radius.  The left panel shows $\theta$-integrated  total mass of new stars within each
given radial bins.  { Note that in our grid configuration, the bin size increases as the
radius (see \S~\ref{sec:setup}).} We can see that for rotating galaxies there are two peaks
for each curve. The outer peaks are purely due to the geometry effect, as explained in our
\citetalias{Yuan:18}. The inner peaks are absent in the case of non-rotating galaxy and become
present due to the rotation of the galaxy.  From model k03 to k05, with the increase of the galaxy
rotation, the amplitude of the peaks increases and the corresponding radius of the peaks becomes
larger: the peak value of the model k05 (mild rotator) is an order of magnitude larger than that
of the model k03 (slow rotator).  To understand these results, we note that the peak location
corresponds to the edge of mid-plane disk shown in Figure~\ref{fig:dnewst}.  The overall star
formation density in the disk is very high, since star formation is strong there. This explains
the presence of the peaks and why the location of the peak of the model k03 is smaller.
But interestingly, both the magnitude and location of the peaks
``saturate'' from model k05 to k09.

For the model without AGN feedback, k05noFB, the star formation occurs actively within 10
pc. We can see from the left panel of Figure~\ref{fig:dnewst} that the mass of the new stars
in this model is up to two orders of magnitudes larger than the models with AGN feedback. It
is because the surface density of the disk is high due to the lack of disturbance by wind and
radiation from the AGN. If the entire disk were stable over the evolution time, the total mass
of new stars in k05noFB would be larger than that in the other models. However, as discussed,
the violent stellar feedback occurs intermittently, which expels large amounts of gas, reducing
the total mass of new stars at $r>10$ pc.

The right panel of Figure~\ref{fig:newst} shows the enclosed mass of the newly born stars within a
given radius at the end of the run. It is clear that as the galaxy rotates faster, the total mass of
new stars becomes larger: the total masses of the newly born stars for the model k00 and k05 are
$6.5\times 10^{9}\, M_{\odot}$ and $1.9\times 10^{10}\, M_{\odot}$, which correspond to 2\% and 6\%
of the initial galactic stellar mass, respectively.  It is because higher angular momentum leads the
gas to stay longer at the mid-plane disk, inducing more active star formation.

\begin{figure}[!htbp]
    \centering
    \includegraphics[width=0.5\textwidth]{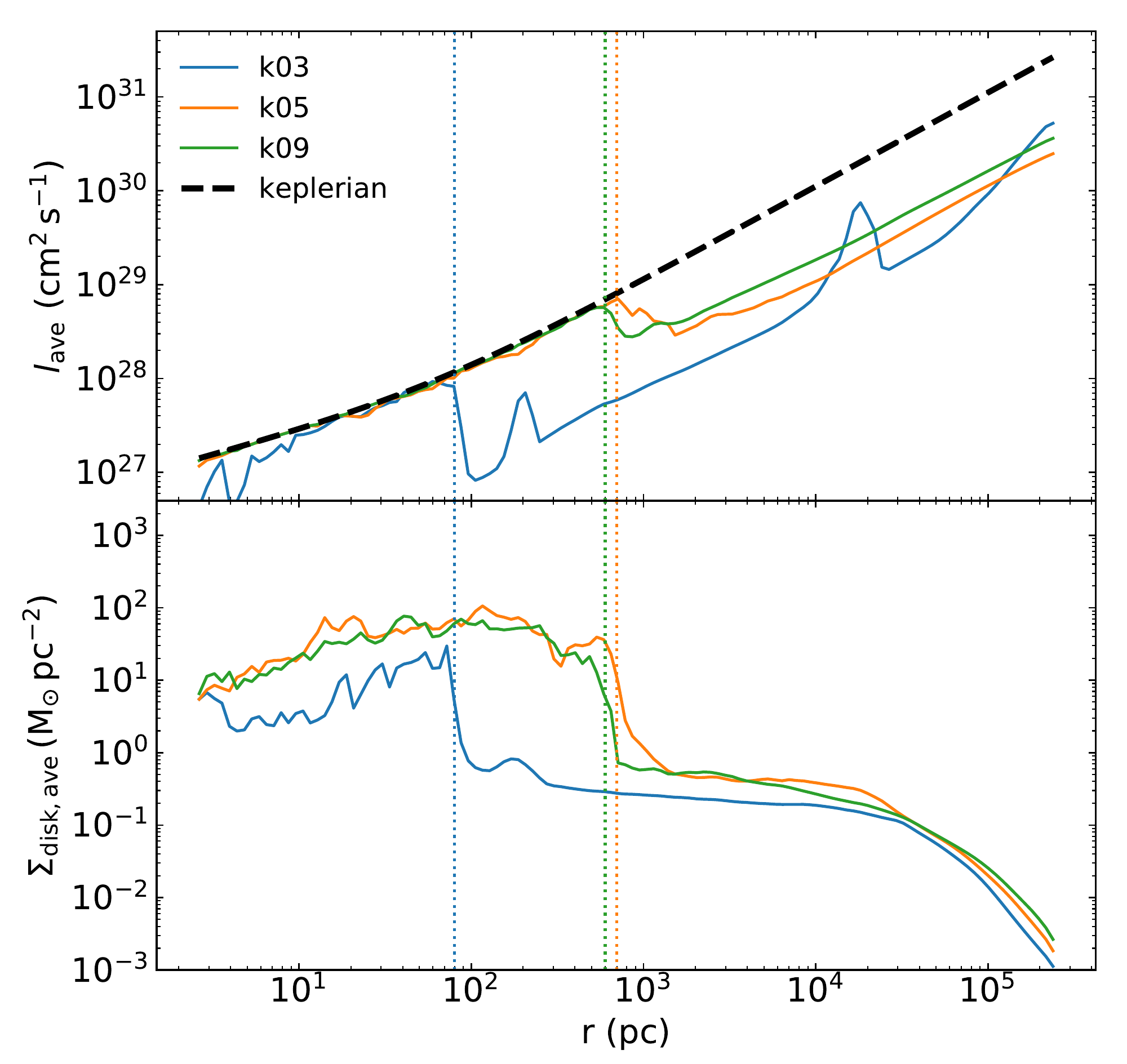}
    \caption{The time-averaged specific angular momentum ($l_{\rm ave}$;upper panel) and surface
            density ($\Sigma_{\rm disk,ave}$;lower panel) of the mid-plane disk as a function of
            radius. The integrated time is from 2 Gyr to 3 Gyr. The specific angular momentum is
            density-weighted, and the disk is identified within $\sim 15^{\circ}$ above/below
            the equator.  The black dashed line represents the  Keplerian value with
            the given gravitational potential by the black hole and dark matter. The vertical dotted lines
            mark the cut-off location where the values drop sharply.}
    \label{fig:disk}
\end{figure}

\begin{figure*}[!htbp]
    \centering
    \includegraphics[width=0.7\textwidth]{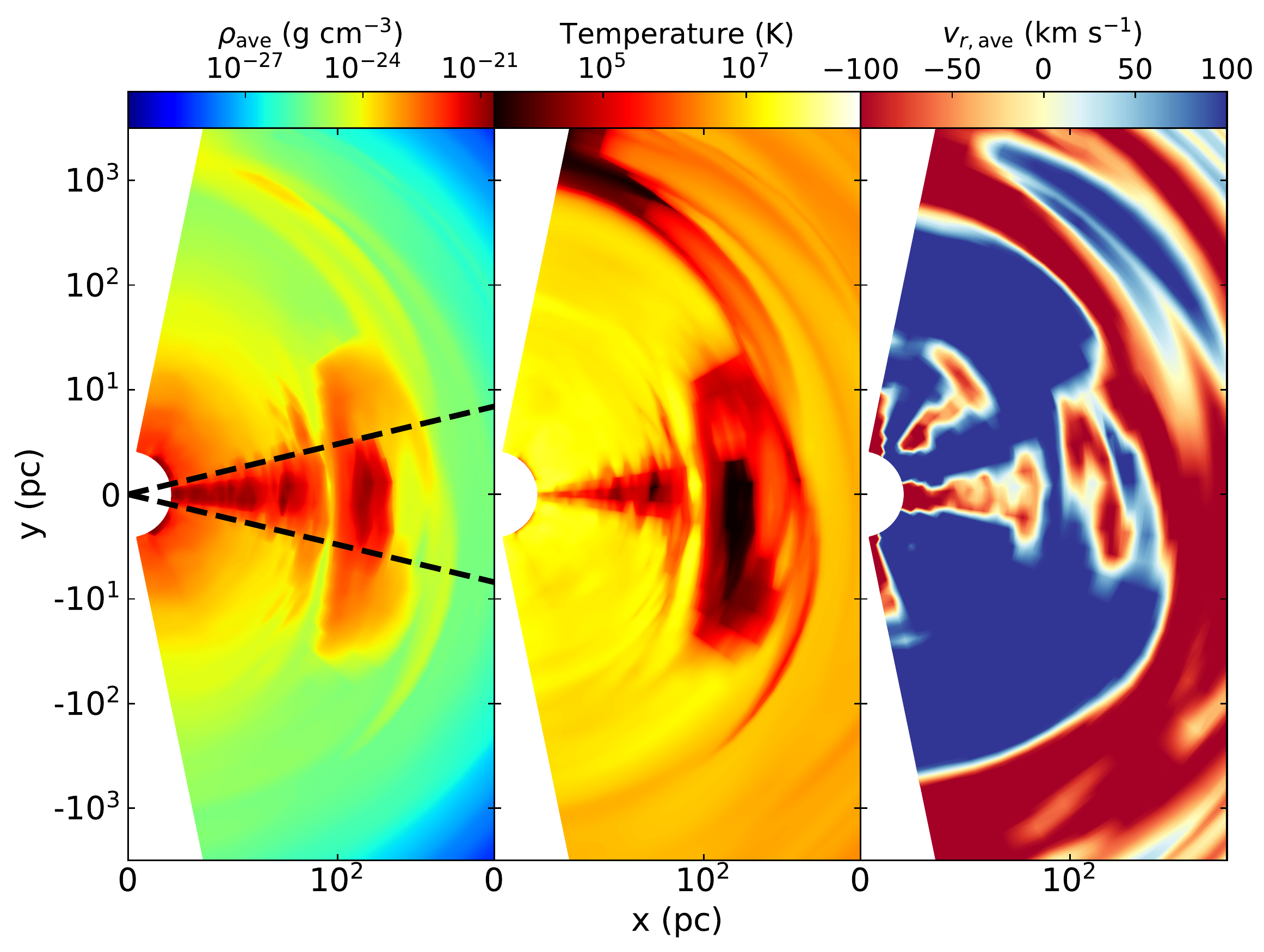}
    \caption{A contour map of the time-averaged density, temperature and radial velocity. The mid-plane disk
            is identified within the black dashed line (left-most panel), which is $15^{\circ}$ above/below the equator. \\~~~~}
    \label{fig:c5show3}
\end{figure*}

The detailed disk properties are illustrated by the radial profiles of the specific angular momentum,
$l_{\rm ave}$, and the disk surface density, $\Sigma_{\rm disk,ave}$. The results are shown
in Figure~\ref{fig:disk}. For the radial profiles, we average the data samples that lie in the
time interval between 2 Gyr and 3 Gyr.  The disk surface density is computed within $15^\circ$
above/below the equator (see the black dashed line in Figure~\ref{fig:c5show3}). Once the gas
falls onto the mid-plane disk, the gas follows the Keplerian motion under the given gravitational
potential. The size of the disk is initially determined by the angular momentum of the gas, but
grows gradually due to the outward transport of angular momentum\citep{Bu:14}. For the slowly
rotating model (k03), the sharp cut-off in the radial profiles occurs at the radius of 80 pc,
which is an order of magnitude smaller than the fast rotators (k05 and k09). However, in our
results, the disk of the galaxies that rotate faster than k05 model are no longer stretched
further, which produces saturation of the star formation activity in the disk for the models
k05 and k09 (see Figure~\ref{fig:newst}).

It is notable that the disk surface density of the model k03 is significantly lower than that
of the fast rotators (k05 and k09). This is mainly because the mid-plane disk in the slowly
rotating galaxy tends to be disrupted easily. We found that in a large fraction of time, the
mid-plane disk disappears in the k03 model, and thus the averaged value of the surface density
for the integrated time interval becomes small. Such a short duration is ascribed to the AGN activity,
which is more violent when the host galaxy rotates slower, as we discussed in \S~\ref{subsec:light}.
As a result, such relatively strong AGN feedback disrupts the disk easily, attenuating the star
formation activities in the disk, which is seen in Figure~\ref{fig:newst}.



\begin{figure*}[!htbp]
    \begin{center}$
        \begin{array}{cc}
            \includegraphics[width=0.5\textwidth]{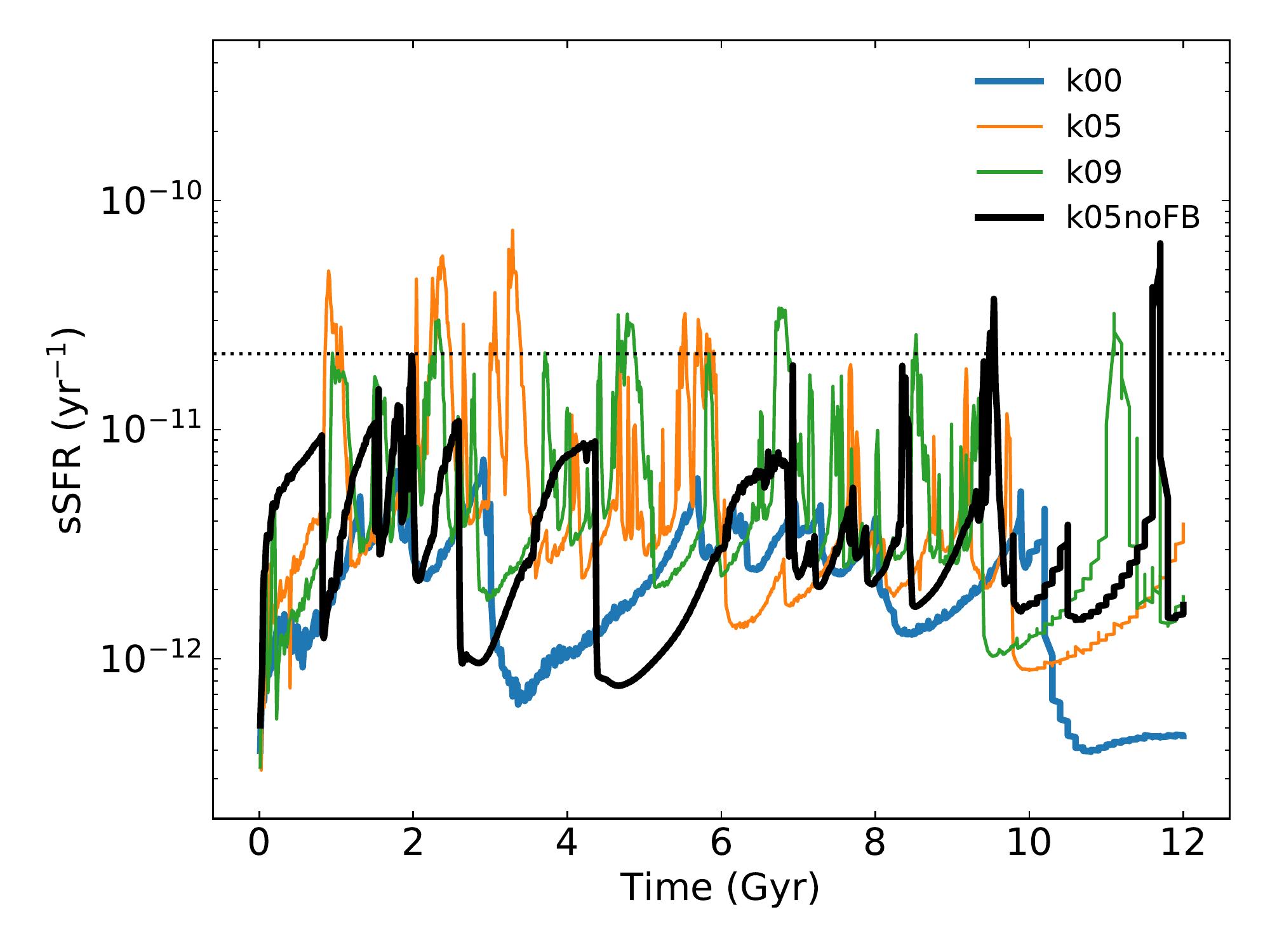} &
            \includegraphics[width=0.5\textwidth]{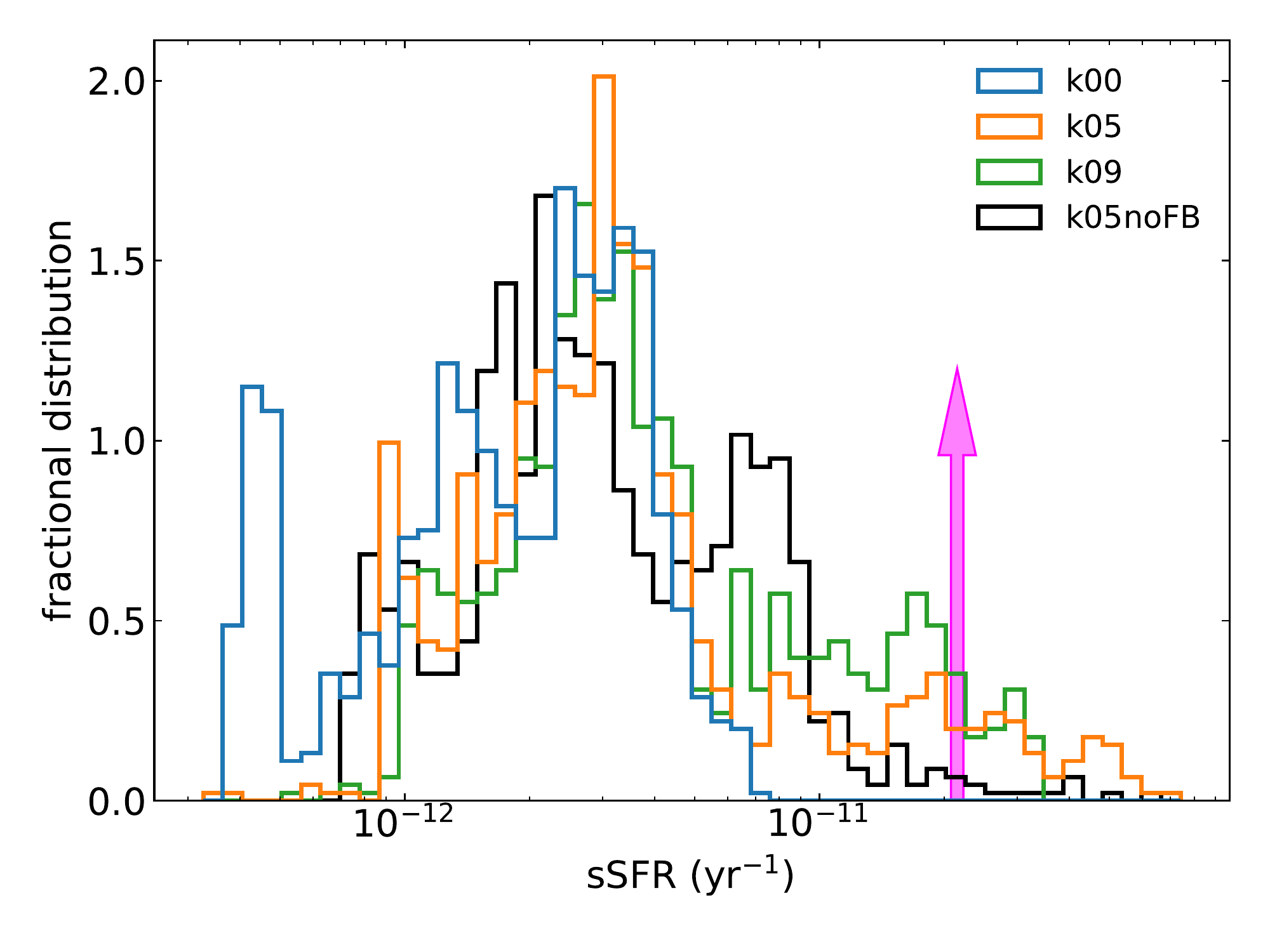}
        \end{array}$
    \end{center}
    \caption{{\it Left panel:} The specific star formation rate as a function of time for the
            models with different galactic rotation.  The horizontal dotted line in the left
            panel and the vertical arrow in the right panel represent the quiescence limit,
            below which the star formation is considered as being quenched \citep{Franx:08}.
            {\it Right panel:} The fractional distribution of the specific star formation rate. The
            colors of histogram, which represent each model, are the same as in the left panel.  \\~~~~}
    \label{fig:sSFR}
\end{figure*}

Since the interplay between the AGN activity and star formation is complicated, to determine whether
the AGN feedback induces or suppresses the star formation, the time evolution of the star formation
rate should be examined.  The left panel of Figure~\ref{fig:sSFR} shows the specific star formation
rate (sSFR) over the galactic evolution, which is computed by star formation rate normalized by the
stellar mass of the galaxy. As discussed in \citetalias{Yuan:18}, the star formation rate is quite
episodic as a consequence of the effects of AGN feedback.  The horizontal dotted line is the
quiescence limit \citep{Franx:08}, below which the star formation is considered as being quenched.
It is interesting to point out that even for the model without AGN feedback, the star formation rate
fluctuates significantly.  Such fluctuation is not by AGN feedback but by both the accretion
process, which is intrinsically fluctuating, and the stellar feedback.  Compared to the model with
AGN feedback, the sSFR in the model without AGN feedback is relatively low. However, this does not
imply that AGN feedback induces star formation: without AGN feedback, stellar feedback becomes
violent intermittently, and expels large amount of gas outward.  In the right panel of
Figure~\ref{fig:sSFR}, the histogram shows that the typical value of the specific star formation
rate for all models is a few $10^{-12} \,\rm yr^{-1}$.

\citet{Smethurst:18} showed observational evidence that, while the rapid quenching of star
formation ($\tau\lesssim$1 Gyr) is dominant for slowly rotating galaxies; the star formation
in fast rotators remains active for a longer evolution time. They argued that it is attributed
to the different nature of the quenching mechanisms: the slowly rotating galaxies may be formed
in major mergers, but the rapidly rotating galaxies are involved in multiple processes, such
as secular evolution and minor mergers.  However, \citet{Lagos:17} showed that the merged
galaxies, which are slowly rotating due to the loss of angular momentum as a consequence of
merging, still need to be quenched by feedback, otherwise the continuing gas inflow and star
formation dominates over the negative effect of mergers.  Our results show that AGN feedback
effectively suppresses star formation activities in slowly rotating galaxies (see the model k00
in Figure~\ref{fig:sSFR}). In rapidly rotating galaxies, the negative effect of AGN feedback
on the star formation activity is relatively weak, as we see that for the models k05 and k09,
the histogram of the specific star formation rate spreads out to the high values.




In order to examine whether the inflowing gas ends up accreting onto the BH
against the depletion by massive star formation along its way, we compare the time scale between
infall, $\tau_{\rm infall}$, and star formation, $\tau_{\rm SF}$:
\begin{equation}\label{eq:infall}
    \tau_{\rm infall} \equiv \frac{r}{v_{r}},
\end{equation}
\begin{equation}\label{eq:SF}
    \tau_{\rm SF} \equiv {\rm max}(\tau_{\rm cool}, \tau_{\rm dyn}),
\end{equation}
where $\tau_{\rm cool}$ is the cooling time scale,
\begin{equation}
    \tau_{\rm cool} \equiv \frac{e}{n^{2}\,\Lambda},
\end{equation}
where $e$ is the internal energy of the gas, $n$ is the number density and
$\Lambda$ is the cooling rate, which is calculated from recent atomic database\footnote{\url{http://atomdb.org/}}.
$\tau_{\rm dyn}$ is the dynamical time scale:
\begin{equation}
    \tau_{\rm dyn} \equiv {\rm min}(\tau_{\rm Jeans},\tau_{\rm rot}),
\end{equation}
where $\tau_{\rm Jeans}$ and $\tau_{\rm rot}$ are Jeans' time scale and rotational time scale, respectively:
\begin{equation}
    \tau_{\rm Jeans} \equiv \left( \frac{3\,\pi}{32\,G\,\rho} \right)^{1/2},
\end{equation}
\begin{equation}
    \tau_{\rm rot} \equiv \frac{2 \pi r}{v_{c}},
\end{equation}
where $v_{c}$ is the Keplerian velocity.

\begin{figure}[!htbp]
    \centering
    \includegraphics[width=0.4\textwidth]{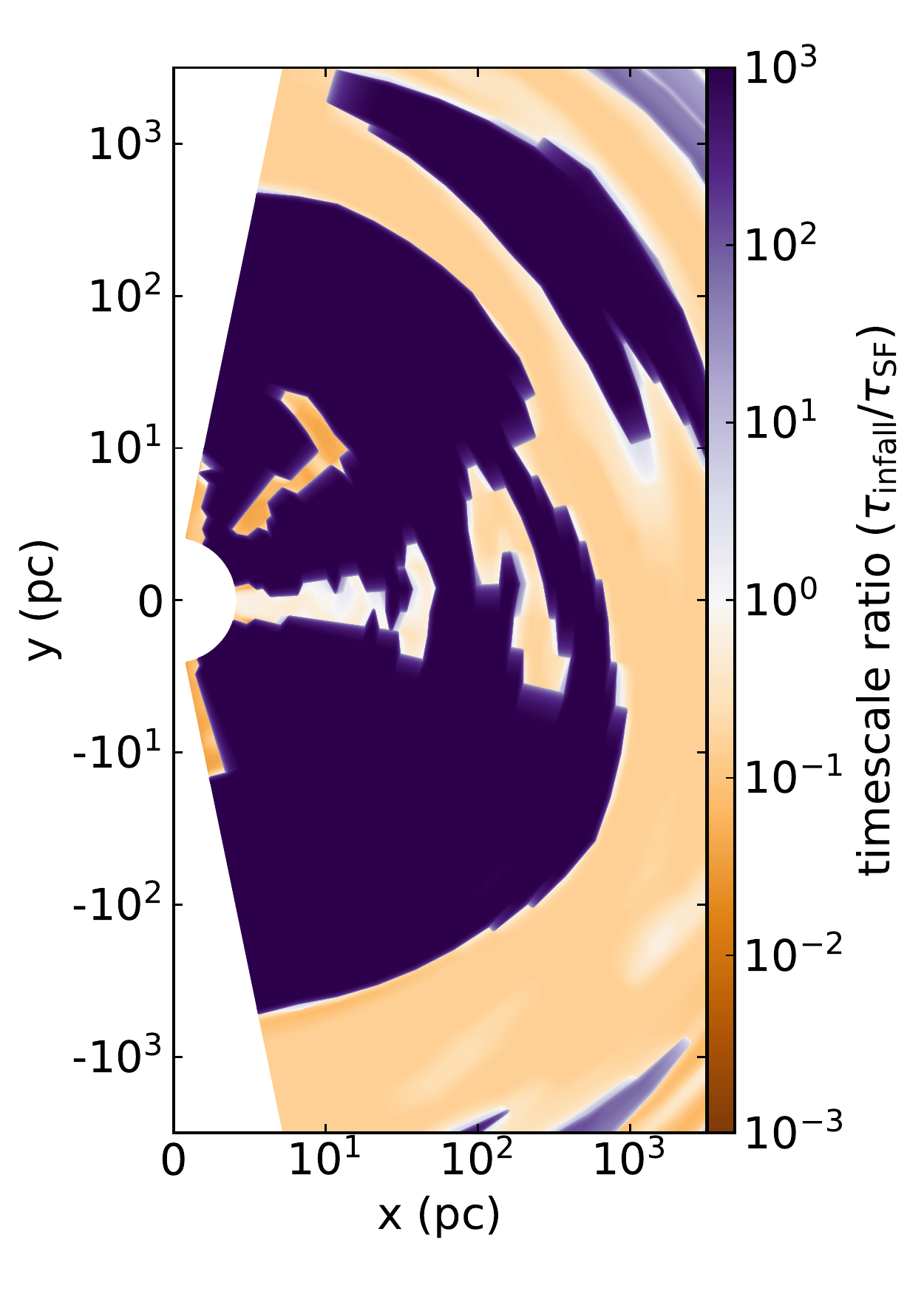}
    \caption{A contour map of the time scale ratio between the infalls and the star formation, $\tau_{\rm infall}/ \tau_{\rm SF}$. \\~~~~}
    \label{fig:tscale}
\end{figure}

To examine different timescales, we need information concerning density, temperature, and radial
velocity. These are calculated and shown in Figure~\ref{fig:c5show3}. The results are  from
the data averaged for some periods, showing the mid-plane cold disk, along which the gas flows
inward with the radial velocity of $\sim -100s$ km/s.  Using this averaged dataset, we examine
the time scale ratio between the infalls and the star formation, $\tau_{\rm infall}/\tau_{\rm
SF}$, from the eqs.~(\ref{eq:infall})\&(\ref{eq:SF}). The results for model k05 are shown in
Figure~\ref{fig:tscale}. From this figure, we find that the infalling time scale is comparable
to the star formation time scale over the entire disk in our fiducial model, k05.  Consequently,
in this model, a large fraction of infalling gas ends up with accreting onto the black hole
against the depletion by star formation, triggering AGN activity as seen in Figure~\ref{fig:ldot}.

It should be noted that, in our simulation, star formation is computed by $\dot{\rho_{\rm
SF} = \eta_{\rm SF}\rho/\tau_{\rm SF}}$ without consideration of temperature and/or Jean's
mass limiters. Here $\eta_{\rm SF}$ is the star formation efficiency and we adopt a value of
$\eta_{\rm SF}=0.1$. The absence of the limiters may lead to an overestimate of star formation
in high temperature and low density regions (e.g., the regions at large radii, $r\gtrsim 10s
\,\rm kpc$). We examined that this does not affect our results significantly, but will be taken
into account in future work.

~~~~

\subsection{AGN Duty Cycle}

\begin{figure*}[!htbp]
    \begin{center}$
        \begin{array}{cc}
            \includegraphics[width=0.5\textwidth]{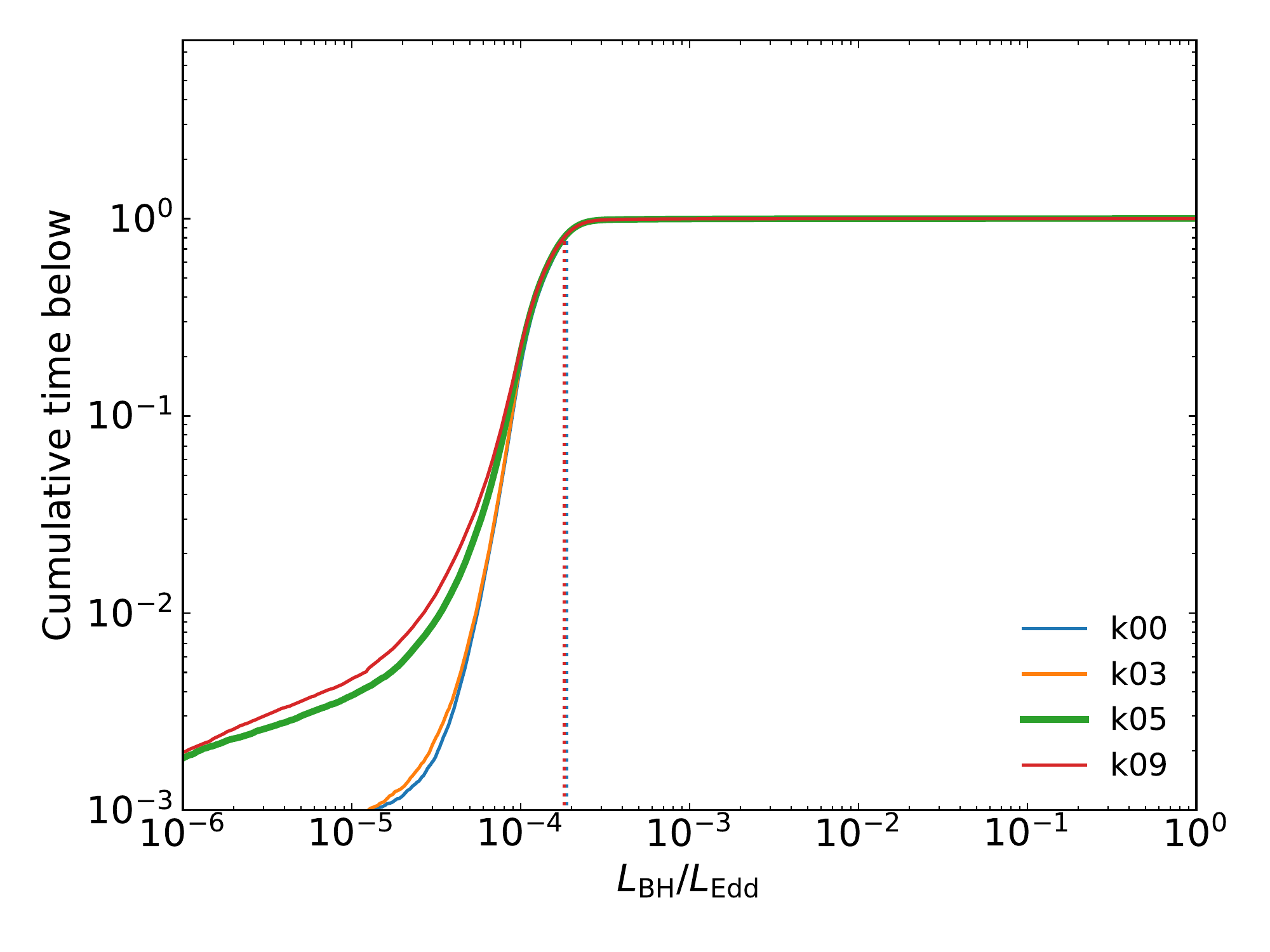} &
            \includegraphics[width=0.5\textwidth]{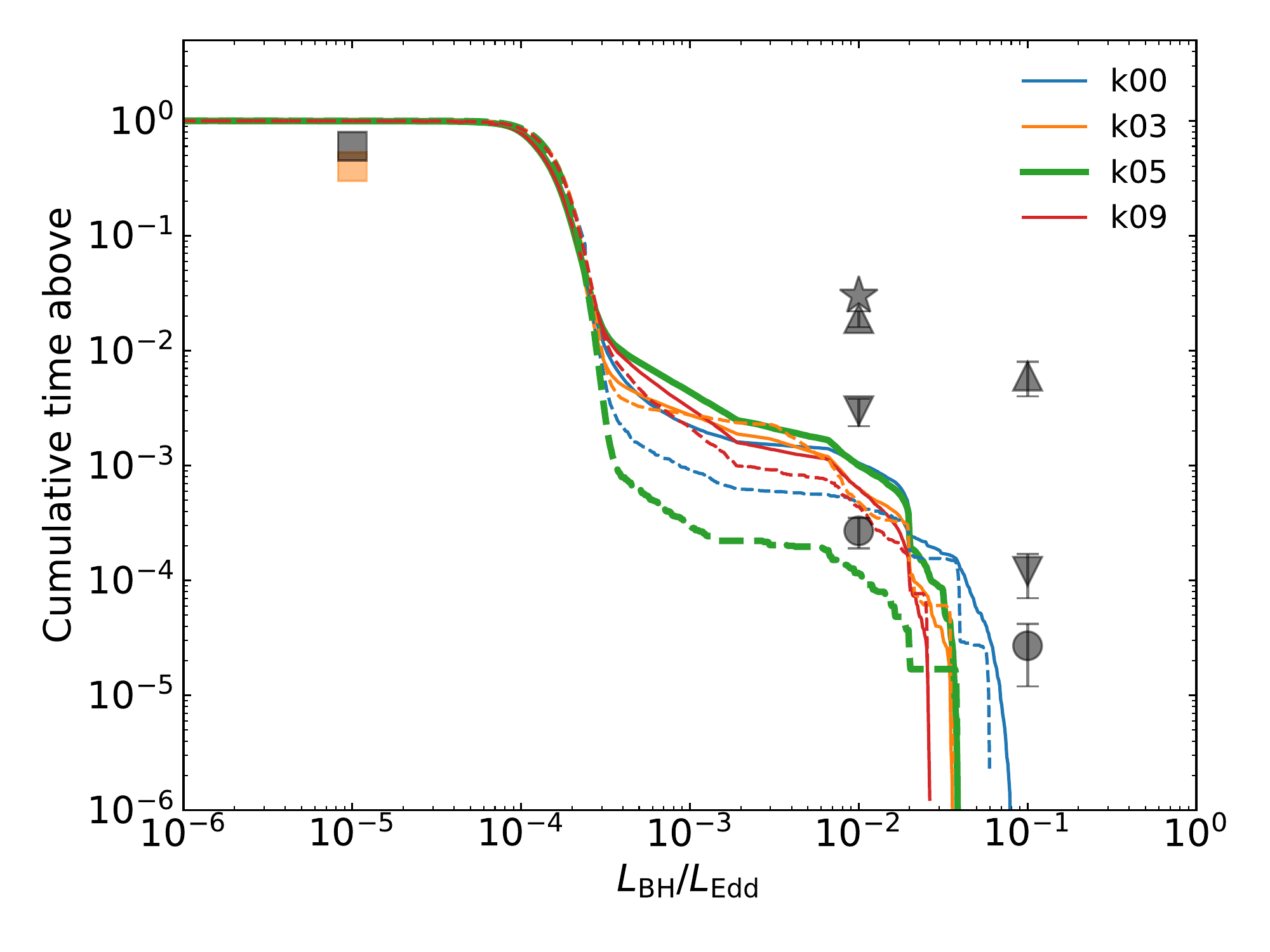}
        \end{array}$
    \end{center}
    \caption{Percentage of the total simulation time spent below (left panel) and above (right
            panel) the given Eddington ratios.  In the left panel, vertical dotted lines indicate
            the Eddington ratio below which the AGN spends 80\% of the total time.  In the
            right panel, the solid and dashed lines represent fractional time computed for the
            entire time and the last 2 Gyrs, respectively. The symbols represent observational
            data point [square:\citet{Ho:09}, circle:\citet{Greene:07}, upwarding-pointing
            triangles:\citet{Kauffmann:09}, downward-pointing triangles:\citet{Heckman:04},
            star:\citet{Steidel:03}]. \\~~~~~}
    \label{fig:duty_t}
\end{figure*}

Following the procedure illustrated in \citet{Ciotti:17b} for the low-rotation case,
Figure~\ref{fig:duty_t} shows the percentage of the total simulation time spent above (right
panel) and below (left panel) the given Eddington ratio. In left panel, the vertical dotted
lines indicate the Eddington ratio below which the AGN spends 80\% of the
total time.  Every model spends most of the time at a low accretion regime (hot mode), which
is consistent with the low detectability of active galaxies \citep[e.g,][]{Greene:07}.

In the right panel of Figure~\ref{fig:duty_t}, the solid and dashed lines represent duty cycle
profiles for the entire evolution time and the last 2 Gyrs, respectively, during which the
AGN spend time above the given Eddington ratio. The observational constraints are indicated by
symbols: downward-pointing triangles are from \citet{Heckman:04}, circles from \citet{Greene:07},
upward-pointing triangles from \citet{Kauffmann:09}, and squares are from \citet{Ho:09}. Those
data points are appropriate for comparing with the low-redshift results (i.e., dashed lines).
The overall duty cycle is similar for all models, but we note that, as the galaxy rotates faster,
the highest AGN luminosity which the AGN can reach becomes smaller gradually: the maximum AGN
luminosity of k09 model is $L_{\rm BH,max}\approx 0.025 \,L_{\rm Edd}$, which is a half of the
maximum luminosity of k00 model. The results show that all models lie below the observation data
but not by a large factor. In our numerical configuration, we assume that the gas is initially
rarefied in the background medium, ignoring the presence of the gas that inflows from the
intracluster medium or the gas that forms before 2 Gyr. Hence, we speculate that our numerical
result of AGN duty cycle in Figure~\ref{fig:duty_t} likely represents the case of lower limit.

\begin{figure*}[!htbp]
    \begin{center}$
        \begin{array}{cc}
            \includegraphics[width=0.5\textwidth]{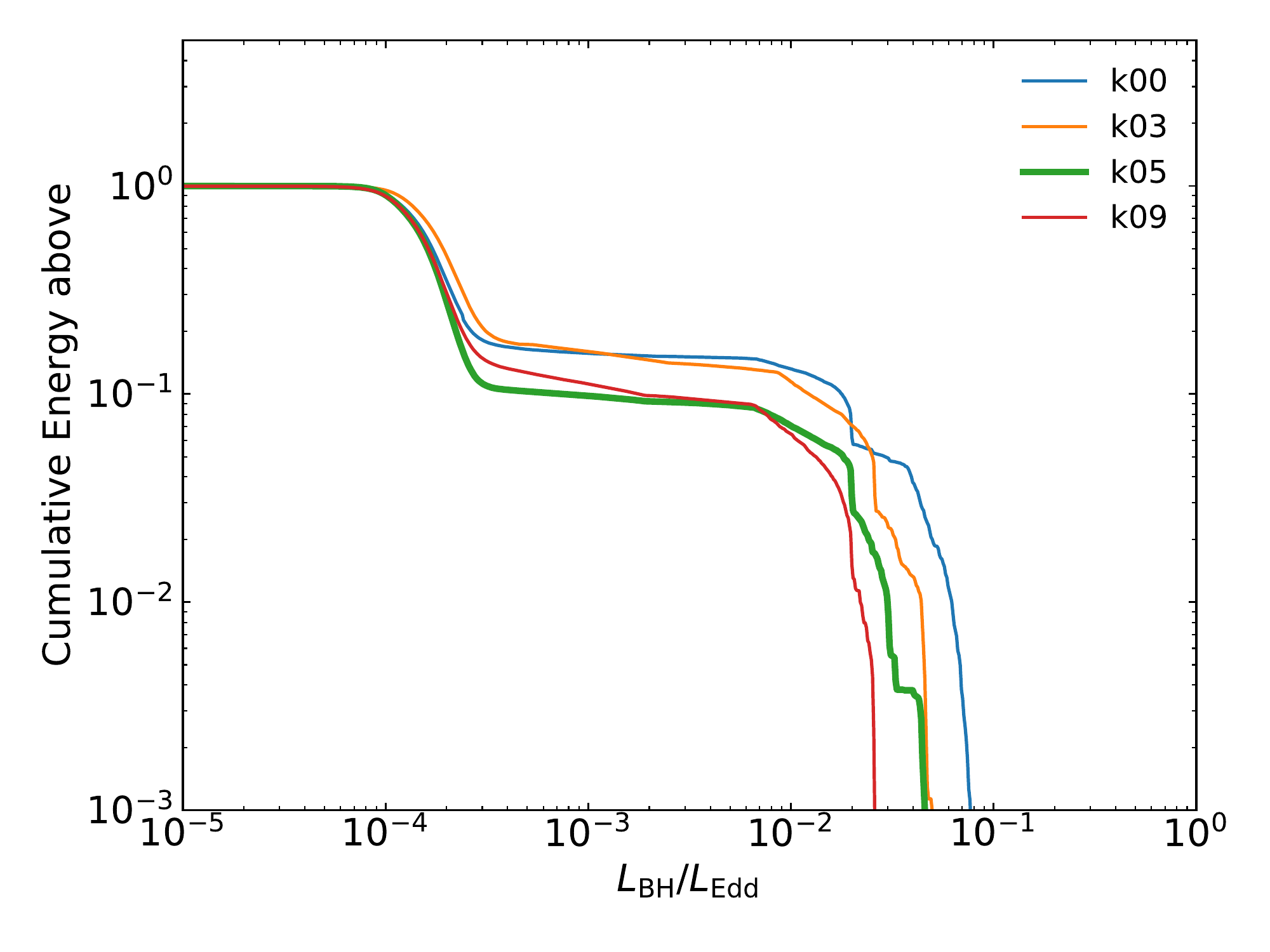} &
            \includegraphics[width=0.5\textwidth]{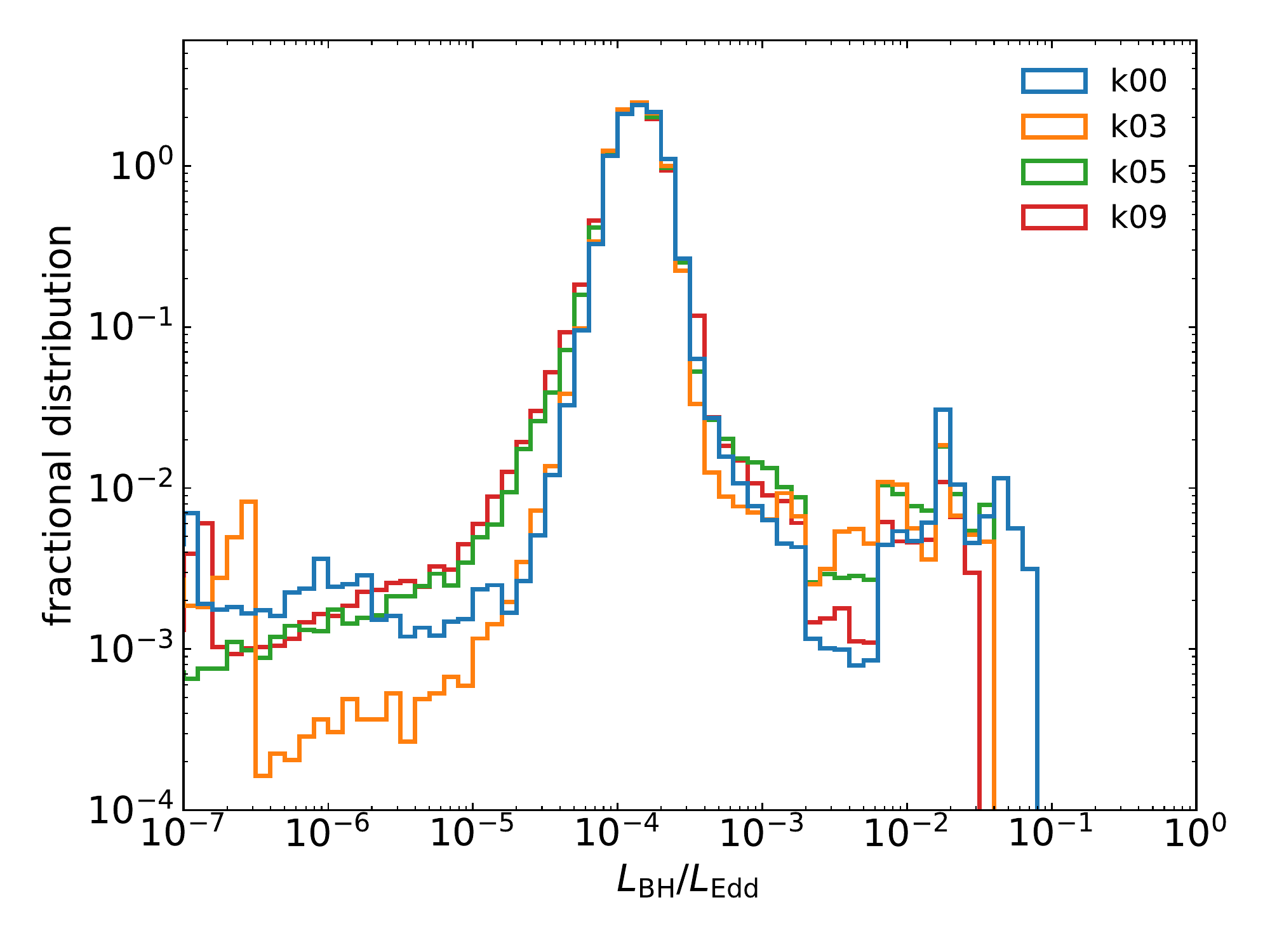}
        \end{array}$
    \end{center}
    \caption{{\it Left panel:} Percentage of the total energy emitted above the values of the
            Eddington ratios.  {\it Right panel:} The fractional distribution of the black
            hole luminosity, which is normalized by the Eddington value.  The colors of the histograms,
            which represent each model, are the same as in the left panel.\\~~~~~}
    \label{fig:duty_L}
\end{figure*}

\citet{Soltan:82}, \citet{Caplar:15} and \citet{Kollmeier:06} argued that while AGNs spend most of
time in the low Eddington regime, they emit significant fraction of energy in the high Eddington
regime.  The left panel of Figure~\ref{fig:duty_L} shows the percentage of the total energy emitted
above the given Eddington ratios. We found that as the galaxy rotates faster, the fraction of
emitted energy with the high Eddington ratio decreases.  The total energy emitted in the cold mode
(i.e., $L_{\rm BH}/L_{\rm Edd} > 0.02$) in the model k09 is several per cents of the entire energy
emitted via AGN feedback, which is an order of magnitude smaller than that in the model k00.  This
trend also can be seen in the fractional distribution of the Eddington ratios (the right panel of
Figure~\ref{fig:duty_L}).  It is clear that the angular momentum plays a role in reducing the
strength of AGN feedback and that all our models emit less energy at high Eddington ratios than do
real observed massive black holes.

~~~~

\subsection{X-ray Properties of the Gas}

We compute X-ray luminosity within the energy band range of 0.3-8 keV (the {\it Chandra
sensitive band}).
\begin{equation}
    L_{\rm X} = 4\,\pi \int^{\infty}_{0} \, \varepsilon \left( r \right) \,r^{2}\,dr,
\end{equation}
where the emissivity is computed by $\varepsilon \left(r \right)=n_{e} \left( r \right) n_{\rm
H}\left( r \right) \Lambda \left[T \left( r \right) \right]$, $n_{e}$ and $n_{\rm H}$ are
the number densities of electrons and hydrogen atoms, and $\Lambda \left( T \right)$ is the cooling
function. We make use of the spectral fitting package with the assumption of the collisional
ionization equilibrium XSPEC\footnote{\url{http://heasarc.gsfc.nasa.gov/docs/xanadu/xspec/}}
(spectral model APEC) to calculate the cooling function \citep{Smith:01}.

\begin{figure*}[!htbp]
    \begin{center}$
        \begin{array}{cc}
            \includegraphics[width=0.5\textwidth]{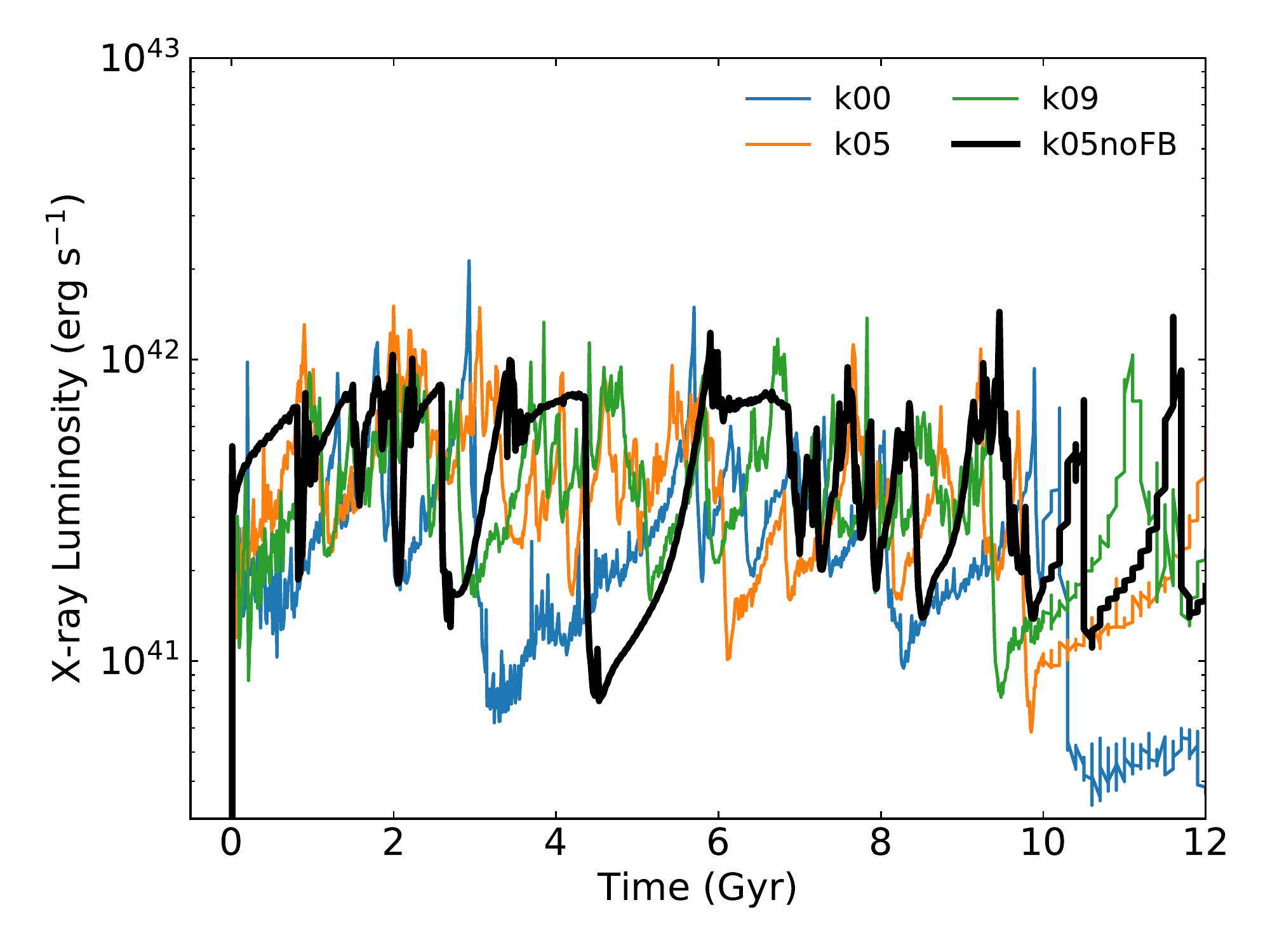} &
            \includegraphics[width=0.5\textwidth]{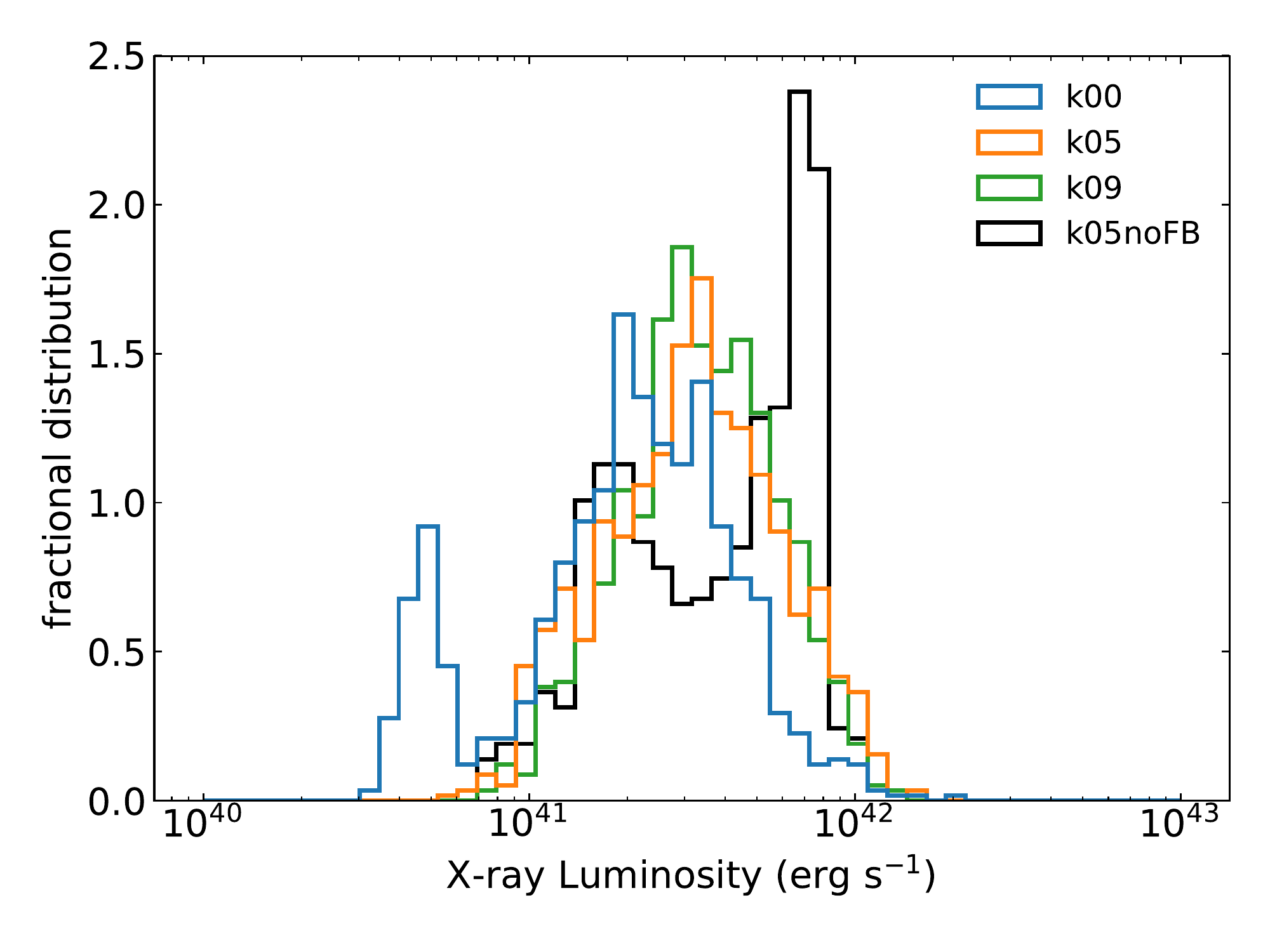}
        \end{array}$
    \end{center}
    \caption{{\it Left panel:} X-ray luminosity in the 0.3-8 keV band over time for the different models.
            {\it Right panel:} The fractional distribution of X-ray luminosity. The colors of histogram,
            which represent each model, are the same as in the left panel. \\~~~~}
    \label{fig:xlum}
\end{figure*}

The X-ray output is mostly emitted from the hot gas in the central region, where the effects of
AGN feedback are likely dominant.  The left panel of Figure~\ref{fig:xlum} shows X-ray luminosity,
$L_{\rm X}$ as a function of evolution time.  We found that for all models with AGN feedback,
the $L_{\rm X}$ strongly oscillates due to the AGN activity, and lies in the range $L_{\rm X}\sim
10^{41}-10^{42}\, \rm erg\,s^{-1}$ in general. For most of models, the dominant X-ray luminosity
band is $\sim 3\times 10^{41}\,\rm erg\,s^{-1}$ (see the right panel of Figure~\ref{fig:xlum}).
As discussed in \citetalias{Yuan:18}, such values are consistent with observations \citep[see
also][]{Pellegrini:18}.

~~~~~~


\section{Summary and Conclusions}\label{sec:summary}

In this work, we have investigated the interplay between the AGN outputs released from the
small scale BH accretion and its host galaxy on a large scale. The primary goal is to
understand the role of the angular momentum of accreting gas in such interplay. We performed
two-dimensional hydrodynamic simulations, in which the spatial range covers from several pc to
$\sim 100s$ kpc.  The galaxy model (stellar distribution and dark matter) and physical processes
(e.g., star formation, Type Ia and Type II supernovae) are described in \citetalias{Yuan:18}.
We adopt the most updated ``sub-grid'' AGN physics as described in detail in \citetalias{Yuan:18},
in which there are two modes of black hole accretion according to the mass accretion rate (i.e.,
hot and cold), and the description of wind and radiation are different in the two modes.

The high angular momentum is a natural barrier for BH accretion due to angular momentum
conservation. Most previous numerical studies of AGN feedback have assumed a galaxy model with
a very low level of angular momentum. However, the restriction is not a good approximation
for most galaxies, even for early type galaxies.  In this work, we remove this restriction
and investigate carefully how the level of angular momentum affects the interplay between AGN
and its host galaxy. In order to transport angular momentum, we adopt the $\alpha$-viscosity
model with an anomalous stress tensor. The fiducial value of the viscosity parameter is set to
$\alpha_{\rm visc}=0.1$ for most models in this work.

The main results are described below:

\begin{itemize}
    \item The general evolution picture of the AGN cycle in a rotating galaxy is qualitatively
        similar to that in a non-rotating galaxy. However, the details are significantly
        different.  The important new feature is the presence of the mid-plane disk for the rotating
        galaxy, due to the angular momentum. As a result, the gas fueling mainly occurs through the
        mid-plane, unlike to the case of non-rotating galaxy, in which the gas fuels the black
        hole in a random direction.
    \item We found that as the galaxy rotates slower, the AGN bursts occur more frequently at
        early evolution times, and the peak of the AGN light curve tends to be higher. The angular
        momentum of the gas plays a role in reducing the AGN activity by two aspects: first,
        the mass accretion rate is reduced due to the presence of angular momentum. Second,
        it forms a mid-plane disk, within which a large fraction of gas is consumed via star
        formation before the gas reaches the BH.
    \item
        The reduced AGN activity in the case of a rapidly rotating galaxy tends to disturb the
        mid-plane disk less strongly than in a slowly rotating galaxy. Consequently, as the host galaxy
        rotates faster, stars form dominantly at the mid-plane disk, and the total mass of new
        stars increases.
    \item The overall profile of duty cycle is similar for all models with different levels of
        angular momentum, and it is somewhat below the observed data. Since we assume that the
        initial gas density is rarefied in the simulation, our results likely represent the
        minimum AGN activity in a given galactic environment. Too little energy is emitted at
        high Eddington ratio compared to observations, which is likely also due to this assumption.
    \item The X-ray luminosity is, in general, similar for all models, and the value is in a
        good agreement with the observed data.
\end{itemize}

~~~~~

\section*{Acknowledgments}
DY and FY are supported in part by the National Key Research and Development Program of
China (Grant No. 2016YFA0400704), the Natural Science Foundation of China (grants 11573051,
11633006, 11650110427, 11661161012),  the Key Research Program of Frontier Sciences of CAS
(No. QYZDJSSW-SYS008), and the Astronomical Big Data Joint Research Center co-founded by
the National Astronomical Observatories, Chinese Academy of Sciences and the Alibaba Cloud.
ZG is supported by the Natural Science Foundation of Shanghai (grant 18ZR1447200).  This work
made use of the High Performance Computing Resource in the Core Facility for Advanced Research
Computing at Shanghai Astronomical Observatory.

~~~~~

\bibliographystyle{aasjournal}
\bibliography{AGNRot}

\end{document}